\documentclass{article}
\pdfoutput=1
\usepackage{youngtab}
\usepackage{lipsum}
\usepackage{enumerate}
\usepackage{amsfonts}
\usepackage{amsmath}
\usepackage{comment}
\usepackage{amsthm, tikz}
\usepackage{amssymb}
\usepackage{color}
\usepackage{graphicx}
\usepackage{pdflscape}
\usepackage[matrix,arrow,curve]{xy}
\usepackage{array}

\usepackage[]{breqn}
\def\K{\mathcal{K}}
\def\nn{\nonumber}
\def\sp{\sigma}

\makeatletter
\renewcommand*{\@cite@ofmt}{\bfseries\hbox}
\makeatother

\hoffset= - 1.2in         

\begin{document}

\title{\vspace{0.1cm}{\Large {\bf Distinguishing Mutant Knots}\vspace{.2cm}}
\author{\ {\bf L. Bishler$^{a,b,c}$}, {\bf Saswati Dhara$^{d}$}, \ {\bf T. Grigoryev$^{e}$}, \ {\bf A. Mironov$^{a,b,c}$}, \ {\bf A. Morozov$^{b,c,e}$},\\  \ {\bf An. Morozov$^{e,b,c}$}, \  {\bf P. Ramadevi$^{d}$}, \ {\bf Vivek  Kumar Singh$^{d}$},\ {\bf A. Sleptsov$^{b,c,e}$}}
\date{ }
}

\maketitle

\vspace{-5.5cm}

\begin{center}
\hfill FIAN/TD-12/20\\
\hfill IITP/TH-09/20\\
\hfill ITEP/TH-12/20\\
\hfill MIPT/TH-09/20\\
\end{center}

\vspace{4.2cm}

\begin{center}

$^a$ {\small {\it Lebedev Physics Institute, Moscow 119991, Russia}}\\
$^b$ {\small {\it ITEP, Moscow 117218, Russia}}\\
$^c$ {\small {\it Institute for Information Transmission Problems, Moscow 127994, Russia}}\\
$^d$ {\small {\it Department of Physics, Indian Institute of Technology Bombay, Mumbai 400076, India}}\\
$^e$ {\small {\it Moscow Institute of Physics and Technology, Dolgoprudny, 141701, Russia }}
\end{center}

\vspace{1cm}

\begin{abstract}
Knot theory is actively studied both by physicists  and mathematicians  as it provides a connecting centerpiece for many physical and mathematical theories. One of the challenging problems in  knot theory is distinguishing mutant knots. Mutant knots are not distinguished by colored HOMFLY-PT polynomials for knots colored by either  symmetric and or antisymmetric representations of $SU(N)$.
Some of the mutant knots can be distinguished by the simplest non-symmetric representation $[2,1]$. However there is a class of mutant knots which require more complex representations like $[4,2]$. In this paper we calculate polynomials and differences for the mutant knot polynomials in representations $[3,1]$ and $[4,2]$ and study their properties.
\end{abstract}


\vspace{.5cm}

\tableofcontents
--
\section{Introduction}

Knot theory is an active area of research at the interface of mathematics and physics. One of the challenging problems attempted from various approaches is classification of knots. One of the ways to approach this problem is to use knot/link invariants, which must coincide for topologically equivalent knots. Unfortunately, some inequivalent knots can share the same knot invariants if the latter are not powerful enough.

In this paper, we will confine to the construction of colored HOMFLY-PT polynomials which are generalisations of the well-known HOMFLY-PT polynomials \cite{HOMFLY,PT}. These generalised invariants are obtained when we place higher dimensional representation of $SU(N)$ as colors on knots. The construction of these knot polynomials have been attempted from 3d Chern-Simons theory \cite{CS} as Wilson-loop averages which is intimately connected to Wess-Zumino conformal field theory \cite{Witt}.
These invariants involve monodromy braiding eigenvalues and the fusion matrices relating conformal blocks \cite{RTmod1,RTmod12,RTmod13,NRZ}. The HOMFLY-PT invariants are written as polynomial in two variables
$q=\exp [2 i \pi / (k+N)]$, $ A=q^N$ where $k$ is the Chern-Simons coupling constant, which is also the level of the
$SU(N)_k$ Wess-Zumino conformal field theory.
Finding the polynomial form of these generalised invariants for arbitrary $SU(N)$ representation is still an open problem as the fusion matrices are not known\footnote{An algorithmic procedure of evaluating the HOMFLY-PT polynomial colored with an arbitrary representation is known so far only for torus knots: it is the celebrated Rosso-Jones formula \cite{RJ1,RJ2} obtained within a completely different approach. It allows one to move on in many directions \cite{BEMT1}-\cite{Dunin} for this class of knots.}. These fusion matrices are known for symmetric, antisymmetric and some low dimensional  non-rectangular representation allowing the computation of the corresponding colored HOMFLY-PT polynomials. Using these polynomials, the reformulated invariants of many knots in topological string theory \cite{GV1,GV2,OV,LMVm,LMV,LM,TS1} could be explicitly worked out \cite{LMOV} and shown to obey the Ooguri-Vafa conjectured form. Other theories related to the knot theory are integrable systems, both classical \cite{ModernRT2,Int1,Int2} and quantum \cite{akutsu,ADO}. This is not that surprising: the classical integrability is related to character expansion of knot invariants \cite{ModernRT1}, while the quantum one, to the Reshetikhin-Turaev construction of knot invariants \cite{RT}-\cite{RT5} and use quantum $\mathcal{R}$-matrices. This connection also makes a bridge to quantum computations \cite{QC1}-\cite{QC3}.


 There is a series of knots which are not distinguished by the colored HOMFLY-PT polynomials when any symmetric or antisymmetric representations are placed on the knots. For instance, a special procedure involving a $180^\circ$ rotation of a 2-tangle (called \textit{mutation}) in a knot $\mathcal K$  can change the knot to $\mathcal K_m$. This operation does not change colored HOMFLY-PT polynomials in (anti)-symmetric and rectangular representations, i.e. they coincide for $\mathcal K$ and $\mathcal K_m$. Actually, the mutation operation can be shown to be a trivial (\textit {identity}) operation for the representations $T$ such that every irreducible representations $Q_i$'s in the tensor product
 \begin{equation}
 T \otimes T = \oplus_i Q_i
 \end{equation}
 occurs only once (multiplicity equal to one). Hence, in order to have a non-trivial action of the mutation operation, one needs to deal with non-rectangular representations which allow some irreducible representations $Q_i$  to occur more than once (with multiplicity higher than 1). In practice, some of the mutant knots are distinguished by the HOMFLY-PT polynomials in representation $[2,1]$, the smallest non-rectangular representation \cite{Mor,Rama1}. However, there are some mutant knots which are not distinguished by the representation $[2,1]$ and require at least representation $[4,2]$ \cite{Mor2}: the increase in the dimension of the multiplicity subspace is an underlying reason for distinguishing such mutants knots.

 Our aim is to compute  the polynomial form of $[4,2]$ colored  HOMFLY-PT invariants for these mutant pairs. There are different approaches to calculating colored HOMFLY-PT polynomials. The most effective ones  are  based on the papers \cite{RT}-\cite{RT5} and use quantum $\mathcal{R}$-matrices . 
 Amongst these, there are  approaches  involving  braid representation of a knot is  \cite{ModernRT1,ModernRT2}, \cite{IMMM3}-\cite{ModernRTfin}. These have an advantage of transforming $\mathcal{R}$-matrix into a very simple diagonal form, but require calculations of the Racah coefficients which are proportional to the fusion matrices of Wess-Zumino conformal field theory.  Note that the number of required Racah coefficients rises with both size of a representation and number of strands in a braid. Another similar approach allows one to calculate colored HOMFLY-PT polynomials from the fundamental HOMFLY-PT polynomials of a cabled knot \cite{Cab}. The approach from Chern-Simons theory related to Wess-Zumino conformal field theory involves  different type of Racah coefficients (fusion matrices) and allows one to calculate polynomials for two-bridge and arborescent knots \cite{Rama1},\cite{NRZ}-\cite{RTmod13},\cite{GKR}-\cite{mut21}. Unfortunately, at the moment, none of these approaches allow us to calculate representation $[4,2]$ HOMFLY-polynomials for the mutant knots.

Hence, in  the present paper, we use the Reshetikhin-Turaev (RT) approach to calculate knot polynomials as a product of $\mathcal{R}$-matrices for a particular quantum group $U_q(sl_N)$ where the deformation parameter is the parameter of the polynomials $q=\exp[2i\pi/(k+N)]$. We use this approach to find differences between HOMFLY-PT polynomials of the mutant knots and study their properties.

Previously, only differences between the polynomials of mutant knots for some particular quantum group $U_q(sl_N)$ and for some particular knots were known, that is, the differences for representation $[2,1]$ of the group $U_q(sl_3)$ for the pretzel mutant knot pair\footnote{The knot notation can be found, e.g., in \cite{barnatan}.  For enumeration of mutants with up to 16 intersections see \cite{M1,M2}.} $K11n34-K11n42$ (the famous Kinoshita-Terasaka and Conway knots) \cite{Mor} and for representation $[4,2]$ of the group $U_q(sl_3)$ for the pretzel mutant knot pairs (in accordance with notation of \cite{arborcalc2}) $K(3,3,3,-3,-3)-K(3,3,-3,3,-3)$ and $K(1,3,3,-3,-3)-K(1,3,-3,3,-3)$ \cite{Mor2}. Recently, using the arborescent knots approach \cite{mut21}, we managed to systematically calculate the answers for the HOMFLY polynomials of arborescent mutant knots in representation $[2,1]$. This allowed us to study the structure and the dependence on $N$ of the differences between the polynomial invariants of mutant knots\footnote{The resulting differences are also reproduced in the present paper in eqn.(\ref{mutlist}).}. However, higher representations were still not studied.

Using the RT approach in the present paper, we have managed to systematically construct differences between the HOMFLY polynomials of mutant knots in representation $[3,1]$. This allowed us to study their structure and compare it to the differences in representation $[2,1]$. We also checked that these colored HOMFLY polynomials passed the standard checks like the factorization property at $q=1$ \cite{DMMSS,IMMMfe}, etc. Unfortunately, a recently realized property of the HOMFLY-PT polynomials in the hook representations \cite{My} is non-trivial for higher representations only.

One of the important properties of HOMFLY-PT polynomials, which is reflected in these differences, is differential expansion, first introduced for knot polynomials themselves \cite{DGR},\cite{IMMMfe},\cite{GS}-\cite{Diff3}. The differential expansion for the differences between polynomials of mutant knots, including those calculated in this paper are studied in more details in \cite{Mila1}. Also the RT approach allowed us to find the invariants in representation $[4,2]$ of the $U_q(sl_3)$ and $U_q(sl_4)$ groups for different sets of mutants.

\section{RT approach \label{RT}}

There are several methods of calculating HOMFLY-PT polynomials. The most effective ones involve quantum $\mathcal{R}$-matrices, we will consider this type of approaches. Quantum $\mathcal{R}$-matrices, used in this approach are known from the theory of quantum groups $U_q(sl_N)$ \cite{FRT,qD}.  Thus, we start with reminding the notation in this theory and describe the quantized universal enveloping algebra of $sl_N$.

\subsection{Quantum groups\label{Sqg}}

The  quantized universal enveloping algebra is generated by the elements $E_i$, $F_i$ and $K_i = q^{h_i}$ and $K_i^{-1}$, which satisfy the following commutation relations:
\begin{equation}
\begin{array}{llll}
[h_i, h_j] = 0,& [h_i, E_j] = a_{ij} E_j, & [h_i, F_j] = -a_{ij} F_j, & [E_i, F_j] = \delta_{ij} \frac{q^{h_i}-q^{-h_i}}{q-q^{-1}},\\
\end{array}
\end{equation}
where $a$ is the Cartan matrix of $sl_N$
\begin{equation}
a_{ij}=\left\{\begin{array}{l}
\phantom{-}2,\ \ \ i=j \\
-1,\ \ \ i=j\pm 1 \\
\phantom{-}0,\ \ \ \text{otherwise}.
\end{array}\right.
\end{equation}
 In the theory of $sl_N$, the number of different generators is: $\# h_i = N-1$ and $\#E_i =\# F_i = N(N-1)/2$. Further, these generators satisfy the Serre relations:
\begin{equation}
\begin{array}{l}
E_i E_i E_{i+1}\ \ \ - (q+q^{-1}) E_i E_{i+1} E_i\ \ \  + E_{i+1} E_i E_i\ \ \ = 0, \\
E_{i+1} E_i E_{i+1} - (q+q^{-1}) E_{i+1} E_i E_{i+1} + E_i E_{i+1} E_{i+1} = 0, \\
F_i F_i F_{i+1}\ \ \ \ - (q+q^{-1}) F_i F_{i+1} F_i\ \ \ \ + F_{i+1} F_i F_i\ \ \ \ = 0, \\
F_{i+1} F_i F_{i+1}\  - (q+q^{-1}) F_{i+1} F_i F_{i+1}\ + F_i F_{i+1} F_{i+1}\  = 0. \\
\end{array}
\end{equation}
Another defining property of the quantum group is the quantized coproduct:
\begin{equation}
\begin{array}{l}
\Delta(E_i)=E_i\otimes q^{h_i} + 1\otimes E_i, \\
\Delta(F_i)=F_i\otimes 1 + q^{-h_i}\otimes F_i, \\
\Delta(h_i)=h_i\otimes 1 + 1\otimes h_i.
\end{array}
\end{equation}
All of these properties together allow one to study the representation theory of $U_q(sl_N)$ quantum group, and it is very similar to the representation theory of the corresponding Lie group. In fact, it has the same set of representations at $|q|\ne 1$, but various quantities like the dimensions of irreducible representation, 3-j and 6-j symbols etc. are replaced with their quantum counterparts. This gives rise to replacing ordinary positive integers $n$ by the quantum numbers $[n]_q$
\begin{equation}
[n]_q=\frac{q^n-q^{-n}}{q-q^{-1}}.
\end{equation}
Similar to quantum physics leading to classical physics when $\hbar \rightarrow 0$, one can check that the quantum number $[n]_q  \rightarrow n$ when $q=e^{\hbar} \rightarrow 1$.
\subsection{$\mathcal{R}$-matrix}

\begin{figure}[h!]
  \centering
\begin{picture}(90,60)(150,0)

\qbezier(30,40)(15,30)(30,20)

\qbezier(30,40)(40,45)(50,40)
\qbezier(30,20)(45,15)(59,26)

\qbezier(50,40)(65,30)(70,20)
\qbezier(65,33)(67,35)(70,40)

\put(40,42){\vector(-1,0){2}}

\put(40,18){\vector(1,0){2}}

\put(78,28){\mbox{$=$}}
\put(93,50){\line(0,-1){40}}
\put(93,10){\vector(0,1){25}}


\qbezier(136,44)(134,46)(130,50)

\qbezier(136,16)(134,14)(130,10)
\qbezier(150,10)(120,30)(150,50)

\qbezier(142,20)(150,30)(142,40)

\put(150,50){\vector(1,1){2}}
\put(130,50){\vector(-1,1){2}}


\put(163,28){\mbox{$=$}}

\put(185,50){\line(0,-1){40}}
\put(185,10){\vector(0,1){25}}

\put(200,50){\line(0,-1){40}}
\put(200,10){\vector(0,1){25}}


\put(250,10){\line(1,1){40}}
\put(250,50){\line(1,-1){17}}
\put(273,27){\line(1,-1){17}}

\put(257,53){\line(0,-1){7}}
\put(257,40){\line(0,-1){20}}
\put(257,14){\line(0,-1){7}}

\put(257,30){\vector(0,1){2}}
\put(280,40){\vector(1,1){2}}
\put(280,20){\vector(1,-1){2}}

\put(300,28){\mbox{$=$}}

\put(320,10){\line(1,1){40}}
\put(320,50){\line(1,-1){17}}
\put(343,27){\line(1,-1){17}}

\put(353,53){\line(0,-1){7}}
\put(353,40){\line(0,-1){20}}
\put(353,14){\line(0,-1){7}}

\put(353,30){\vector(0,1){2}}
\put(330,40){\vector(1,-1){2}}
\put(330,20){\vector(1,1){2}}
\label{rand}
\end{picture}
  \caption{Reidemeister moves}\label{FReid}
\end{figure}
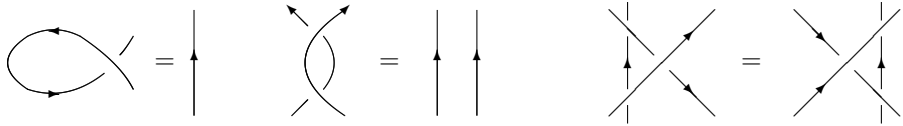

The knot invariant associated with any knot diagram (projection of the knot on a plane) does not change under the three Reidemeister moves (see Fig.\ref{FReid}). The operators corresponding to crossings in the knot diagram
are the braid generators. It is easy to understand that the third Reidemeister move is the familiar quantum Yang-Baxter equation satisfied by the braid generators. The words constructed using braid generators constitute braid group. Clearly, the braid group is not a free group because the braid generators obey defining relations. One can construct the braid generators using the main object in the RT approach, which is the quantum $\mathcal{R}$-matrix.
Hence, knot invariants can be constructed using the quantum $\mathcal{R}$-matrix.

The quantum $\mathcal{R}$-matrix itself can be constructed starting from the universal $\mathcal{R}$-matrix \cite{FRT}:
\begin{equation}
\mathcal{R} =q^{\sum_{i,j} a^{-1}_{i,j} h_i \otimes h_j} \overrightarrow{\prod_{\beta \in \Phi^{+}}} {\rm exp}_{q}     \left ( (q - q^{-1}) E_{\beta} \otimes F_{\beta}   \right),
\end{equation}
In fact, the $\mathcal{R}$-matrix that is used in the braid group is additionally multiplied by the permutation operator $\mathcal{P}$,
\begin{equation}
\mathcal{P} (x \otimes y) = y \otimes x,
\end{equation}
$\Phi^{+}$ are all positive roots of the algebra and the quantum exponential is defined as
\begin{equation}
{\rm exp}_q X = \sum_{m = 0}^{\infty} \frac{X^m}{[m]_q !} q^{m(m-1)/2}.
\end{equation}

In the concrete representations, the $\mathcal{R}$-matrix acts on the product of linear spaces of two representations, $T_1$ and $T_2$ of $U_q(sl_N)$. Since the $\mathcal{R}$-matrix commutes with the coproduct \cite{FRT}, it acts trivially by multiplying with an eigenvalue on the whole irreducible representation $Q$ in the tensor product of $T_1$ and $T_2$:
\begin{equation}
\{\lambda_{\mathcal{R}}\}=\bigcup_{Q\vdash T_1\otimes T_2} \lambda_Q
\end{equation}
This eigenvalue is equal to
\begin{equation}
\label{Reig}
|\lambda_Q|=q^{\varkappa_Q},\ \ \ \varkappa_Q=\sum_{i,j\in Q} (i-j)
\end{equation}
where $i$ and $j$ numbers refer accordingly to the row and the column in the Young diagram associated with $Q$.
The signs of the eigenvalues are defined when $T_1=T_2=T$ by whether the representation $Q$ belong to the positive or negative square of $T$ and are undefined otherwise. Basically, they can be calculated by putting $q=1$ and looking at the sign of the permutation operator eigenvalues. The detailed explanation of how to find these eigenvalues can be found in \cite{mut21,MMMS21,MultiLink}.

\subsection{Modern RT approach for braids \label{MRT}}

The RT approach is well-developed 
for diagrams of knots in the form of a closure of a braid (see Fig.\ref{Fbraid41}). We call this type of approach the modern RT approach. Let us review the main ideas of this method.
\begin{figure}[h!]
\begin{picture}(100,100)(-130,-20)
\put(0,0){\line(1,0){30}}
\put(30,0){\line(1,1){25}}
\put(0,25){\line(1,0){30}}
\put(0,50){\line(1,0){80}}
\put(30,25){\line(1,-1){10}}
\put(55,0){\line(-1,1){10}}
\put(55,25){\line(1,0){25}}
\put(55,0){\line(1,0){25}}
\put(80,50){\line(1,-1){25}}
\put(80,25){\line(1,1){10}}
\put(105,50){\line(-1,-1){10}}
\put(80,0){\line(1,0){50}}
\put(105,25){\line(1,0){25}}
\put(105,50){\line(1,0){75}}
\put(130,0){\line(1,1){25}}
\put(130,25){\line(1,-1){10}}
\put(155,0){\line(-1,1){10}}
\put(155,0){\line(1,0){80}}
\put(155,25){\line(1,0){25}}
\put(180,50){\line(1,-1){25}}
\put(180,25){\line(1,1){10}}
\put(205,50){\line(-1,-1){10}}
\put(205,25){\line(1,0){30}}
\put(205,50){\line(1,0){30}}
\end{picture}
\caption{Braid representation for figure-eight knot: $(a_{1,1}\equiv a_1=1, b_{1,2}\equiv b_1=-1,a_{2,1}\equiv a_2 =1,b_{2,2}\equiv b_2=-1)$}
\label{Fbraid41}
\end{figure}

To calculate the HOMFLY-PT polynomials colored with representation $R$, one needs to color each strand of braid with this representation. If a diagram of knot contains $m$ strands, one needs $m-1$ different braiding generators  $\mathcal{R}_i$. Each of them acts non-trivially only on representations corresponding to the $i$-th and the $i+1$-th strands:
\begin{equation}
\mathcal{R}_i = \overbrace{\,I\, \otimes \dots \otimes\, I\,}^{\mbox{$i-1$}} \otimes\,\mathcal{R}\,\otimes\overbrace{\,I\, \otimes \dots \otimes\, I\,}^{\mbox{$m-i-1$}}.
\end{equation}
The colored HOMFLY-PT polynomial can be calculated as a weighted trace of product of the matrices corresponding to each crossing in the knot diagram
\begin{equation}
\mathcal{H}_R^{\mathcal{B}} = {\rm Tr}_q \left( \prod_{i=1}^n\prod_{j=1}^{m-1}\mathcal{R}_j^{a_{i,j}}\right).
\label{mbraid}
\end{equation}
Here, $\{a_{i,j}\} \equiv \{a_{i,1},b_{i,2}, \ldots k_{i,j}, \ldots l_{i,m-1} \}$ carries the braiding information of the $m$-strand braid. The braid notation is explained in an example in Fig.\ref{Fbraid41}.
The weighted trace ${\rm Tr}_q$ correspond to the closure of braid and corresponds to insertion of an additional operator, usually denoted by $q^{h^{\otimes m}}=K^{\otimes m}$ (see section\ref{Sqg} for the definitions), into the product of $\mathcal{R}$-matrices. It is needed for the answer to satisfy the first Reidemeister move (see Fig.\ref{FReid}) which changes the number of strands in the braid. This weighted trace also satisfies the condition
\begin{equation}
{\rm Tr}_q^Q (\,I\,) = {\rm dim}_q(Q),
\end{equation}
where ${\rm dim}_q(Q)$ is the quantum dimension of representation $Q$, which is equal to the Schur polynomial $\chi_Q\{p_k\}$ evaluated at the special point $p_k=[Nk]/[k]$.

Let us look at the example of three-strand braid in detail. To calculate the HOMFLY-PT polynomials of three-strand braid $(a_1, b_1 | a_2, b_2| \dots|a_n,b_n)$, one needs two $\mathcal{R}$-matrices: $\mathcal{R}_1$ and $\mathcal{R}_2$
\begin{equation}
\mathcal{H}_R^{(a_1, b_1 | a_2, b_2| \dots|a_n,b_n)} = {\rm Tr}_q \left( \mathcal{R}_1^{a_1}\mathcal{R}_2^{b_1} \mathcal{R}_1^{a_2}\mathcal{R}_2^{b_2} \dots \mathcal{R}_1^{a_n}\mathcal{R}_2^{b_n}  \right).
\end{equation}
Let us work in the basis of  irreducible representations $Q_i$ from decomposition $T\otimes T\otimes T = \oplus_i Q_i^{\oplus k_i}$ ($k_i$ here are multiplicities of the representations $Q_i$ in the product of $T$). As was explained, the $\mathcal{R}$-matrices act trivially, just by multiplying with eigenvalues, in this basis, i.e. they can be reduced to blocks $\mathcal{R}_{i,Q_i}$ of the size $k_i$ in this basis. The expression for the HOMFLY-PT polynomial (\ref{mbraid}) then transforms into
\begin{equation}
\mathcal{H}_R^{\mathcal{B}} = \sum_{Q\in R^{\otimes m}} C_{R,Q}^{\mathcal{B}} \chi_Q.
\end{equation}
The coefficients $C_{R,Q}^{\mathcal{B}}$ depend on the eigenvalues of $\mathcal{R}$-matrix and on elements of the Racah matrices. By definition, the Racah matrices $\mathcal{U}$ describe the deviation from the associativity in the product of representations
\begin{equation}\label{RacahDef}
\mathcal{U}_{T_1T_2T_3}^{Q}: \,\,\, ((T_1\otimes T_2)\otimes T_3 \rightarrow Q) \longrightarrow (T_1\otimes(T_2\otimes T_3) \rightarrow Q).
\end{equation}
The matrices $\mathcal{R}_i$ are connected with each other by the Racah matrices. If one diagonalizes one of them (usually the $\mathcal{R}_1$-matrix), all the others can be calculated from its eigenvalues using the Racah matrices. Thus for the three-strand braids the answer would look like

\begin{equation}
C_{R,Q}^{(a_1, b_1 | a_2, b_2| \dots|a_n,b_n)} = {\rm Tr} \left(  \mathcal{R}_Q^{a_1} \mathcal{U}_Q \mathcal{R}_Q^{b_1} \mathcal{U}_Q^{\dagger} \mathcal{R}_Q^{a_2} \mathcal{U}_Q \mathcal{R}_Q^{b_2} \mathcal{U}_Q^{\dagger} \dots \mathcal{R}_Q^{a_n} \mathcal{U}_Q \mathcal{R}_Q^{b_n} \mathcal{U}_Q^{\dagger} \right).
\end{equation}

The most difficult part in the described method is evaluating the Racah matrices. Unfortunately, existing methods of calculations do not allow us to calculate them for representation [4,2], which we are most interested in. What makes the situation more difficult is that the mutant knots have more than three strands, and more strands require more and more complicated Racah matrices.


\begin{figure}[h!]
\centering
\begin{center}
\includegraphics[width=9cm]{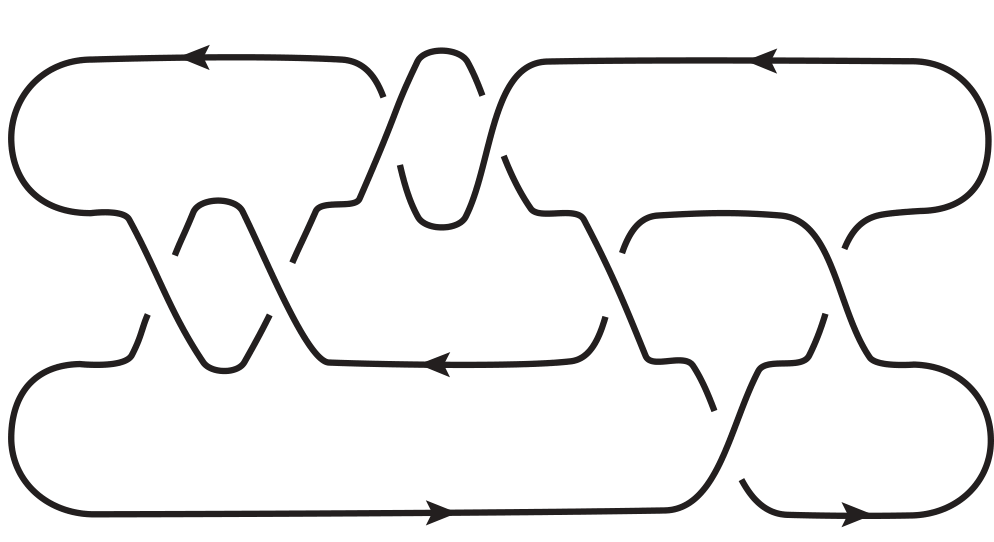}
\end{center}
\caption{Knot $7_6$ as a two-bridge knot.\label{F2br}}
\end{figure}

\subsection{Two-bridge and arborescent knots\label{MRTA}}

Other types of knots for which the modern RT-approach works quite well are the two-bridge knots and their generalization to the arborescent knots \cite{Con,Cau,BS}. Let us start from the two-bridge knots. The two-bridge knots can be represented as a four-strand braid but with a different closure: the strands at each end of the braid are closed with the strands at the same end instead of the opposite end of the braid, see Fig.\ref{F2br}. This means that the structure of representations is different from that in the previous section. For the knot colored with representation $T$, the two strands with arrows pointing right in Fig.\ref{F2br} carry the representation $T$, and the other two carry the conjugate representation $\bar{T}$. The conjugate representation is defined by the property that decomposition of the product of $T$ and $\bar{T}$ includes among other representations the trivial representation,
\begin{equation}
T\otimes \bar{T} = \emptyset \oplus\ldots.
\end{equation}
The closure of the braid used for the two-bridge knot means that one picks up from the (quartic) product of four representations only the trivial representation $\emptyset$. In its turn, this means that which representation should be taken from the sub-product of three representation in this quartic product is determined by the fourth representation. In another words, this means that one actually deals with the same three strand-case as in previous section with the Racah coefficients being a particular case of (\ref{RacahDef}). In fact, there are just three Racah coefficients which are needed for the two-bridge knots:
\begin{equation}\label{RacahDef1}
\begin{array}{lcl}
S: & \, & ((T\otimes T)\otimes \bar{T}\rightarrow T) \longrightarrow (T\otimes(T\otimes \bar{T}) \rightarrow T),
\\
S^{\dagger}: & \, & ((\bar{T}\otimes T)\otimes T\rightarrow T) \longrightarrow (\bar{T}\otimes(T\otimes T) \rightarrow T),
\\
\bar{S}: & \, & ((T\otimes \bar{T})\otimes T\rightarrow T) \longrightarrow (T\otimes(\bar{T}\otimes T) \rightarrow T).
\end{array}
\end{equation}
All other Racah coefficients needed in this case can be found by conjugating all the representations, which does not change the matrix itself. Also these Racah matrices satisfy
\begin{equation}
S S^{\dagger}=I,\ \ \ \bar{S}\bar{S}=I.
\end{equation}
These Racah matrices are called \textit{exclusive} Racah matrices.

Columns and rows in the Racah and $\mathcal{R}$-matrices correspond to irreducible representations from decomposition of the product of two representations. Two-bridge closure of the braid corresponds to the choice of trivial representation among these. The whole braid corresponds to the matrix $B_{XY}$ which is a product of Racah and $\mathcal{R}$-matrices same as it happens for the braid construction in the previous section.
$X$ and $Y$ enumerate, depending on the structure of the knot, either representations in the product $T\otimes T$ or in the product $T\otimes \bar{T}$.
Then the knot polynomial itself corresponds to the matrix element of the resulting matrix $B_{\emptyset\emptyset}$.

\begin{figure}[h!]
\centering
\begin{center}
\includegraphics[width=6cm]{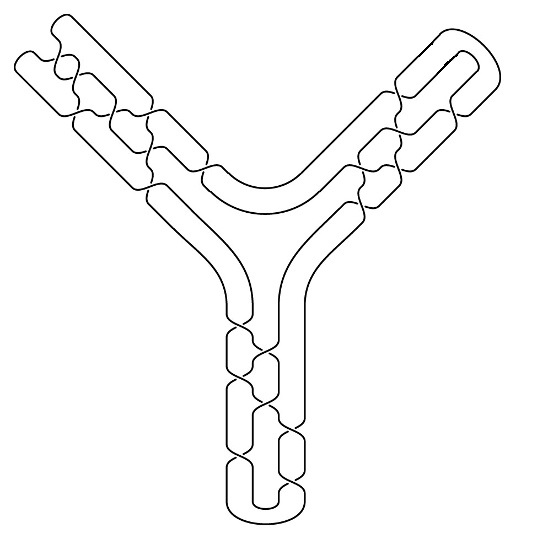}
\end{center}
\caption{Simple arborescent knot which has only leaves and no branches.\label{Farbsimp}}
\end{figure}

This structure can be generalized. If one of the ends of the braid is open, then this structure, finger or leaf corresponds to the operator $B_{X\emptyset}$. If these fingers are connected to each other by pairs of strands (see Fig.\ref{Farbsimp}), then the resulting structure corresponds to the convolution of the corresponding operators giving the knot invariant for Fig.\ref{Farbsimp} as
\begin{equation}
\mathcal{H}\sim\sum_{X \in  { T \otimes \bar T}}  \prod_i^3 B^{(i)}_{X\emptyset}.
\label{star}
\end{equation}

If the both ends are open, one gets the branch which can be connected to other branches or leaves. This gives a structure of arborescent knots. This is a very powerful method. However, in any case, we do not know the exclusive Racah matrices in representations $[4,2]$ and $[3,1]$. Since the exclusive Racah matrices for $[2,1]$ representation\cite{non-rect} was known,  we used the arborescent knot approach to calculate mutant knot polynomials in representation $[2,1]$ in \cite{mut21}.  For other non-rectangular representations, we do not have the exclusive Racah matrices and hence we cannot compute using this approach.

\subsection{Polynomial calculus \label{CRT}}

In order to calculate the HOMFLY-PT polynomials using the RT approach, one has to associate the $\mathcal{R}$-matrix with each crossing on the diagram of knot. The important property of the knot diagram is that one has to choose a (up-down) direction on the plane along which all the crossings should be aligned. The knot strand sometimes changes the direction, which is described by an operator $\mathcal{M}$:
\begin{equation}
\mathcal{M} = q^{h_{\rho}},
\end{equation}
where:
\begin{equation}
\begin{array}{l}
h_{\rho} = \frac{1}{2} \sum_{\alpha \in \Phi^+} h_{\alpha} = \frac{1}{2} \sum_{i = 1}^{N-1} i(N-i) h_i, \\

\end{array}
\end{equation}
Graphically this operator is depicted as a ``hat'' which changes the direction of the line from up to down and vice versa, there are four different types:

\begin{picture}(300,80)(0,-15)
\put(0,40){\line(0,-1){30}}
\put(30,40){\line(0,-1){30}}
\qbezier(0,10)(15,-10)(30,10)
\put(17,0){\vector(1,0){2}}
\put(-5,45){\text{a}}
\put(30,45){\text{b}}
\put(40,20){\text{$=$}}
\put(55,20){\text{$\mathcal{M}^{a}_{b}$}}

\put(125,10){\line(0,1){30}}
\put(155,10){\line(0,1){30}}
\qbezier(125,40)(140,60)(155,40)
\put(142,50){\vector(1,0){2}}
\put(120,0){\text{a}}
\put(155,0){\text{b}}
\put(165,20){\text{$=$}}
\put(180,20){\text{$(\mathcal{M}^{-1})^{a}_{b}$}}

\put(250,10){\line(0,1){30}}
\put(280,10){\line(0,1){30}}
\qbezier(250,40)(265,60)(280,40)
\put(263,50){\vector(-1,0){2}}
\put(245,0){\text{b}}
\put(280,0){\text{a}}
\put(290,20){\text{$=$}}
\put(305,20){\text{$\mathcal{M}^{a}_{b}$}}

\put(375,40){\line(0,-1){30}}
\put(405,40){\line(0,-1){30}}
\qbezier(375,10)(390,-10)(405,10)
\put(388,0){\vector(-1,0){2}}
\put(370,45){\text{b}}
\put(405,45){\text{a}}
\put(415,20){\text{$=$}}
\put(430,20){\text{$(\mathcal{M}^{-1})^{a}_{b}$}}

\end{picture}

In the braid representation of knot, described in section \ref{MRT}, two of these $\mathcal{M}$-operators make a weight matrix $\mathcal{M}^2 = q^{2 h_{\rho}} $, which gives a weighted trace ${\rm Tr}_q$.

Totally, there are eight possible types of crossings. One can express them via the matrices $\mathcal{R}$, $\mathcal{R}^{-1}$, $\mathcal{M}$, $\mathcal{M}^{-1}$ in the following way:

\begin{picture}(500,60)(0,0)
%
\put(40,40){\line(-1,-1){18}}
\put(0,40){\vector(1,-1){40}}
\put(18,18){\vector(-1,-1){18}}
\put(-5,45){\text{a}}
\put(40,45){\text{b}}
\put(-5,-10){\text{c}}
\put(40,-10){\text{d}}
\put(50,20){\text{$=$}}
\put(65,20){\text{$(\mathcal{R}_1)^{ab}_{cd}$}}
\put(100,20){\text{$=$}}
\put(115,20){\text{$\mathcal{R}^{ab}_{cd}$}}
\put(300,40){\vector(-1,-1){40}}
\put(260,40){\line(1,-1){18}}
\put(282,18){\vector(1,-1){18}}
\put(255,45){\text{a}}
\put(300,45){\text{b}}
\put(255,-10){\text{c}}
\put(300,-10){\text{d}}
\put(310,20){\text{$=$}}
\put(325,20){\text{$(\mathcal{R}_5)^{ab}_{cd}$}}
\put(360,20){\text{$=$}}
\put(375,20){\text{$(\mathcal{R}^{-1})^{ab}_{cd}$}}
\end{picture}

\begin{picture}(500,80)(0,0)
%
\put(0,0){\vector(1,1){40}}
\put(0,40){\line(1,-1){18}}
\put(22,18){\vector(1,-1){18}}
\put(-5,45){\text{b}}
\put(40,45){\text{d}}
\put(-5,-10){\text{a}}
\put(40,-10){\text{c}}
\put(50,20){\text{$=$}}
\put(65,20){\text{$(\mathcal{R}_2)^{ab}_{cd}$,}}
\put(105,20){\text{  $\mathcal{R}_2 = (1\otimes \mathcal{M})\mathcal{R}(\mathcal{M}\otimes 1)^{-1}$}}
\put(260,40){\vector(1,-1){40}}
\put(260,0){\line(1,1){18}}
\put(282,22){\vector(1,1){18}}
\put(255,45){\text{b}}
\put(300,45){\text{d}}
\put(255,-10){\text{a}}
\put(300,-10){\text{c}}
\put(310,20){\text{$=$}}
\put(325,20){\text{$(\mathcal{R}_6)^{ab}_{cd}$,}}
\put(365,20){\text{$\mathcal{R}_6 = (1\otimes \mathcal{M}) \mathcal{R}^{-1} (\mathcal{M} \otimes 1)^{-1}$}}
\end{picture}

\begin{picture}(500,80)(0,0)
%
\put(40,40){\vector(-1,-1){40}}
\put(40,0){\line(-1,1){18}}
\put(18,22){\vector(-1,1){18}}
\put(-5,45){\text{c}}
\put(40,45){\text{a}}
\put(-5,-10){\text{d}}
\put(40,-10){\text{b}}
\put(50,20){\text{$=$}}
\put(65,20){\text{$(\mathcal{R}_3)^{ab}_{cd}$,}}
\put(105,20){\text{  $\mathcal{R}_3 = ( \mathcal{M}\otimes 1)^{-1}\mathcal{R}(1 \otimes \mathcal{M})$}}
\put(300,0){\vector(-1,1){40}}
\put(300,40){\line(-1,-1){18}}
\put(278,18){\vector(-1,-1){18}}
\put(255,45){\text{c}}
\put(300,45){\text{a}}
\put(255,-10){\text{d}}
\put(300,-10){\text{b}}
\put(310,20){\text{$=$}}
\put(325,20){\text{$(\mathcal{R}_7)^{ab}_{cd}$,}}
\put(365,20){\text{$\mathcal{R}_7 = (\mathcal{M} \otimes 1)^{-1} \mathcal{R}^{-1} (1 \otimes \mathcal{M})$}}
\end{picture}

\begin{picture}(500,80)(0,0)
%
\put(40,0){\vector(-1,1){40}}
\put(0,0){\line(1,1){18}}
\put(22,22){\vector(1,1){18}}
\put(-5,45){\text{d}}
\put(40,45){\text{c}}
\put(-5,-10){\text{b}}
\put(40,-10){\text{a}}
\put(50,20){\text{$=$}}
\put(65,20){\text{$(\mathcal{R}_4)^{ab}_{cd}$,}}
\put(105,20){\text{  $\mathcal{R}_4 = ( \mathcal{M}\otimes \mathcal{M})\mathcal{R}(\mathcal{M} \otimes \mathcal{M})^{-1}$}}
\put(260,0){\vector(1,1){40}}
\put(300,0){\line(-1,1){18}}
\put(278,22){\vector(-1,1){18}}
\put(255,45){\text{d}}
\put(300,45){\text{c}}
\put(255,-10){\text{b}}
\put(300,-10){\text{a}}
\put(310,20){\text{$=$}}
\put(325,20){\text{$(\mathcal{R}_8)^{ab}_{cd}$,}}
\put(365,20){\text{$\mathcal{R}_8 = (\mathcal{M} \otimes \mathcal{M}) \mathcal{R}^{-1} (\mathcal{M} \otimes \mathcal{M})^{-1}$}}
\end{picture}

\subsection{Multiplicities}

Sometimes, the tensor product $T_1\otimes T_2$ contains the representation $Q$ more than once. This is called multiplicity case. If one looks at formula (\ref{Reig}), it is easy to see that the $\mathcal{R}$-matrix now has two or more eigenvalues which coincide up to a sign. Interestingly, this first happens when we move from symmetric representations to non-symmetric, namely to the representation $[2,1]$. This is the first representation which distinguishes simple mutant knots. Moreover, at this stage another type of coinciding eigenvalues appears. There are pairs of coinciding eigenvalues, while the corresponding representations themselves are different. Detailed studies of these ``accidentally'' coinciding eigenvalues and their properties was done in \cite{BlockR}. Let us provide an example of the structure of eigenvalues:
\begin{equation}
[2,1]\otimes [2,1]=[4,2] \oplus [4,1,1] \oplus [3,3]\oplus 2\cdot [3,2,1]\oplus [3,1,1,1] \oplus [2,2,2] \oplus [2,2,1,1].
\end{equation}
The corresponding eigenvalues (\ref{Reig}) are
\begin{equation}
\begin{array}{cccc}
\lambda_{[4,2]} = \frac{1}{q^5}, &
\lambda_{[4,1,1]} = -\frac{1}{q^3}, &
\lambda_{[3,3]} =  -\frac{1}{q^3}, &
\lambda_{[3,2,1]_{\pm}}  = \pm 1, \\
\lambda_{[3,1,1,1]} = q^3, &
\lambda_{[2,2,2]} = q^3, &
\lambda_{[2,2,1,1]} = -q^5. & \\
\end{array}
\end{equation}
It is easy to see that there are two pairs of accidentally coinciding eigenvalues
\begin{equation}
\begin{array}{c}
\lambda_{[4,1,1]} = \lambda_{[3,3]} =  -\frac{1}{q^3}, \\
\lambda_{[3,1,1,1]} = \lambda_{[2,2,2]} = q^3.
\end{array}
\end{equation}
The representation $[3,2,1]$ has multiplicity two.

However, for the purposes of  our research here, it seems that the eigenvalues that coincide because of multiplicity are more important. We will provide some explanation for this fact in the section \ref{Arb}.

For representation $[2,1]$, these multiplicity eigenvalues coincide only up to a sign. This leads to the fact discovered by Morton \cite{Mor2}, that this representation in fact does not distinguish all possible mutant knots. For some more symmetric mutant knots one needs more complicated representations like $[4,2]$, where the multiplicity $3$ appears ($[4,2]\otimes[4,2]=\ldots\oplus 3\cdot[6,4,2]\oplus\ldots$). This means that two of multiplicity eigenvalues coincide together with the sign.

\section{Mutant Knots}

\begin{figure}[h!]
\centering
\begin{center}
\includegraphics[width=11cm]{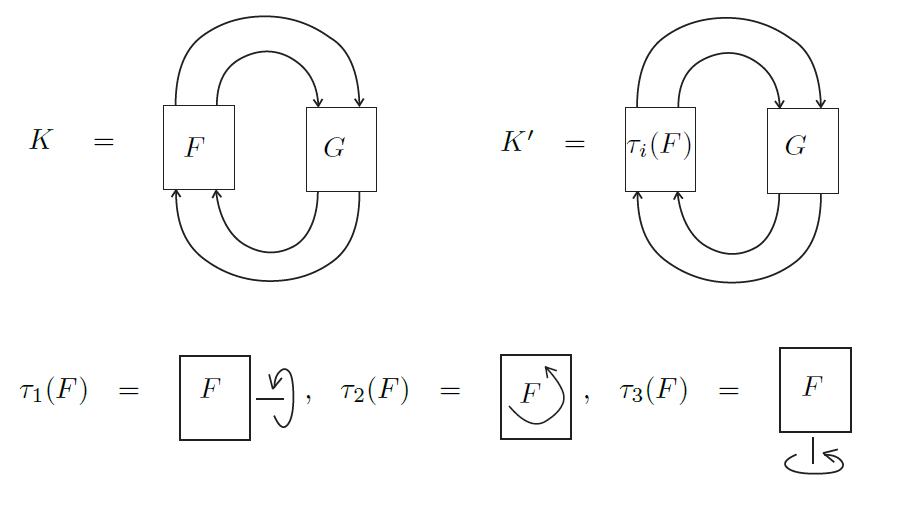}
\end{center}
\caption{Mutation procedure \label{FMut}}
\end{figure}

Mutation is a special procedure which can be applied to the knot that consists of two tangles (see Fig.\ref{FMut}). One of the tangles can be rotated by 180 degrees with respect to some axis and glued back into the knot. If the resulting knot is different from the initial one, these knots are called the \textit{mutant knots}. The interesting property of these mutant knots is that their HOMFLY-PT polynomials in any (anti)symmetric or rectangular representation coincide \cite{Mor}.

Some of the mutant knots, however, can be distinguished by the simplest non-symmetric representation $[2,1]$ \cite{Mor,Rama1}. This happens due to the fact that this representation is the first one where multiplicities appear. An example of mutant knots where the importance of this fact can be comparatively easily understood is the pretzel mutant knots.

\subsection{Pretzel mutant knots \label{Arb}}

The most studied set of knots are torus knots. However there are no mutant pairs among the torus knots. Thus one should look at more complicated sets of knots. One of the straightforward generalizations of the torus knots is pretzel knots. These are the knots that can be placed without self-intersections onto the genus $g$ surface instead of a torus. However, instead of arbitrary number of strands, only two strands are put on each handle. These knots are parameterized by the number of crossings on each handle.

\begin{figure}[h!]
\centering
\begin{center}
\includegraphics[width=5cm]{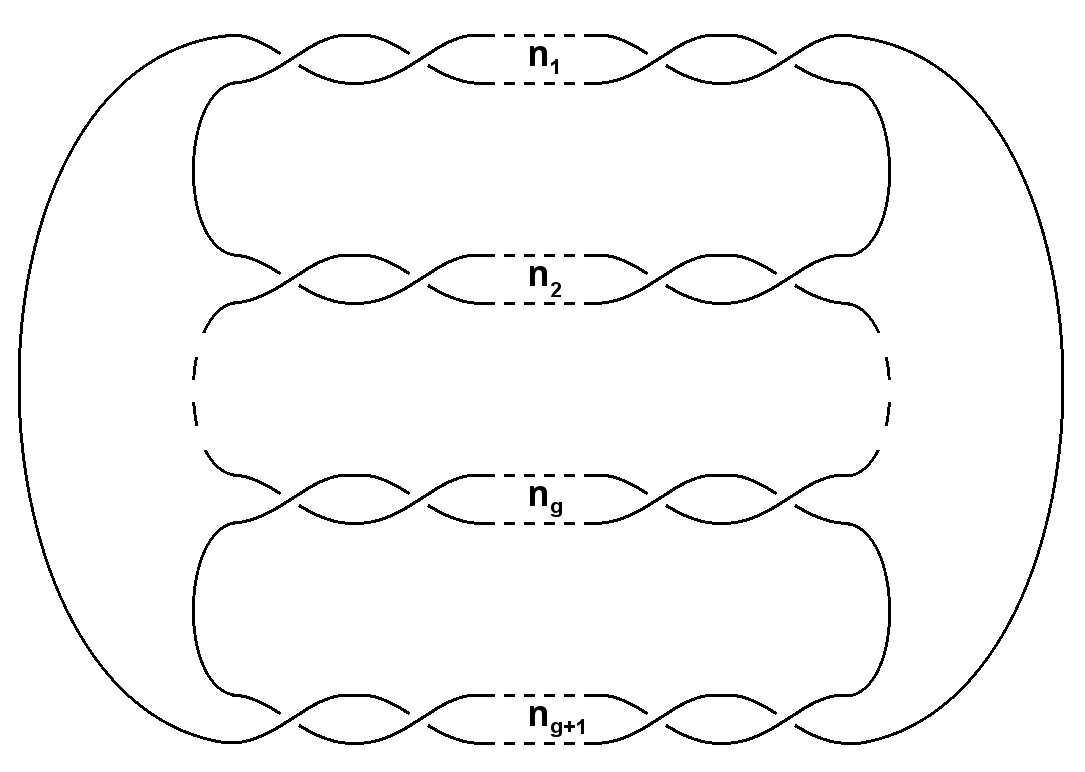}
\end{center}
\caption{Pretzel mutant knot $K(n_1,n_2,\ldots,n_g,n_{g+1})$. Mutation appears if one exchanges numbers $n_i$ and $n_j$. \label{PrKn}}
\end{figure}
These knots are a subclass of arborescent knots. Thus they can be calculated using the same arborescent knot approach (see section \ref{MRTA} and \cite{Rama1,mut21} for details). Like it was done in (\ref{star}), the answer for the knot invariant can be found from
\begin{equation}
\mathcal{H}\sim\sum\limits_X\prod B^{(i)}_{X\emptyset}.
\end{equation}
where $X$ are representations running in the vertex. If there are no multiplicities, each term in the sum is just a product of numbers. Since these numbers commute, the answer would not change if blocks are exchanged, or, for pretzel knots, different braids exchange places. This is exactly a mutation procedure.

However, if $X_i$ has a multiplicity $m_i$, $B_{X_i X_j}$ is a matrix of the size $m_i\times m_i$. These matrices no longer commute, thus the answers for the mutant knots can be different.

\subsection{Morton mutants}

\begin{figure}[h!]
\centering
\includegraphics[width=10cm]{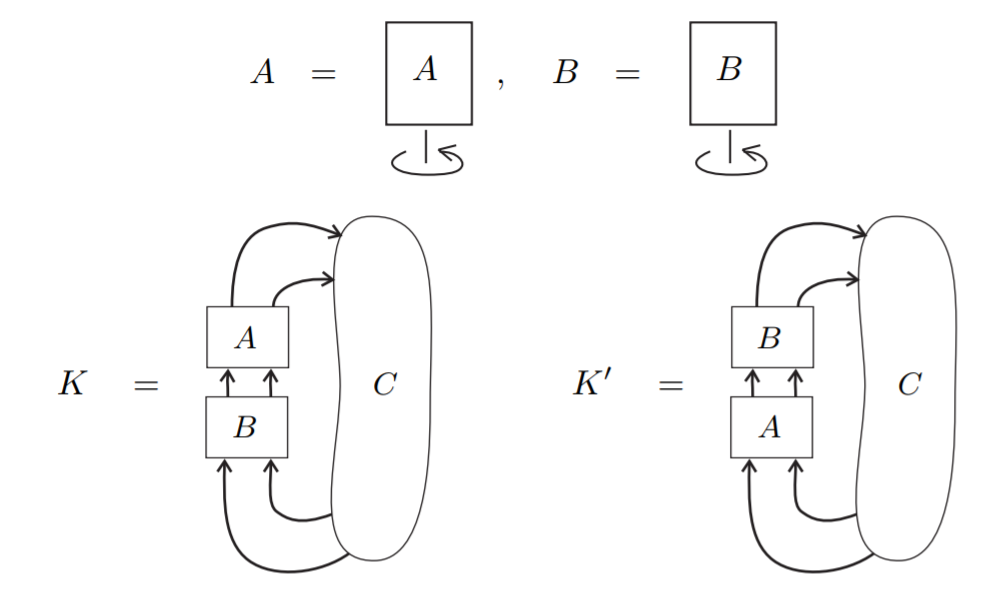}
\caption{Morton mutant knots \label{FMutMor}}
\end{figure}

In \cite{Mor2}, H.Morton suggested that some mutant knots can have a larger symmetry (see Fig.\ref{FMutMor}). Such mutant knots are not distinguished by the HOMFLY-PT polynomials in representation $[2,1]$, unlike simpler mutant knots. The simplest representation that can distinguish them is the representation $[4,2]$. The reason underling this property is the fact that $[4,2]$ is the simplest representation whose tensor product gives an irreducible representation to occur thrice. In other words the multiplicity in this case is $3$.

\section{Mutant knots in representation $[2,1]$ \label{smut11}}

In \cite{mut21}, we have calculated and studied differences between HOMFLY-PT polynomials in representation $[2,1]$ of all 16 pairs of mutant knots with 11 crossings. Since, in this case, the Racah matrices are known, we have used the modern RT approach from Sec.\ref{MRTA} to evaluate these differences. These mutant knots are actually the arborescent knots, which allowed us to use the modern RT approach. We have discovered that, for these knots, the difference has a specific structure
\begin{equation}
\Delta \mathcal{H}_{[2,1]}^{mutant}=A^\gamma\cdot f(A,q)\cdot M(q)
\label{gen21}
\end{equation}
where $\gamma$ is an integer, $M(q)$ is a function of only $q$ (which is a ratio of quantum numbers ) and
$f(A,q)$ \footnote{The explicit expression for $f(A,q)$ here differs from that in \cite{mut21}, since here we use the unreduced (non-normalized) HOMFLY-PT polynomials.} is
\begin{equation}
f(A,q):=\{q\}^{11}\cdot  D_{3}^2D_{2}D_{1} D^2_0 D_{1} D_{-2}D_{-3}^2 [3]^{-1},~\\ \label{ffactor}
\end{equation}
where ``differential" $D_k:=\{Aq^k\}/\{q\}$, and  $\{x\}:=x-x^{-1}$.

These differences of $[2,1]$-colored HOMFLY polynomials for all sixteen 11-crossing mutant pairs are
\begin{equation}\label{mutlist}
\begin{array}{llcl}
1.& \mathcal{H}^{11a44}_{[2,1]} - \mathcal{H}^{11a47}_{[2,1]} &=& A \cdot f(A,q)\cdot\dfrac{[8]}{[2]}\cdot\mathfrak{n}  \\
2.& \mathcal{H}^{11a57}_{[2,1]} - \mathcal{H}^{11a231}_{[2,1]} &=& A^{-5}\cdot f(A,q)\cdot\dfrac{[8]}{[2]}\cdot\mathfrak{n}  \\
3.& \mathcal{H}^{11n71}_{[2,1]} - \mathcal{H}^{11n75}_{[2,1]} &=& A^{13} \cdot f(A,q)\cdot\dfrac{[7][8]}{[14]}\cdot\mathfrak{n}   \\
4.& \mathcal{H}^{11n73}_{[2,1]} - \mathcal{H}^{11n74}_{[2,1]} &=& A^{-3}\cdot f(A,q)\cdot\dfrac{[8]}{[2]}\cdot\mathfrak{n}    \\
5.& \mathcal{H}^{11n76}_{[2,1]} - \mathcal{H}^{11n78}_{[2,1]} &=& A^{-15} \cdot f(A,q)\cdot\dfrac{[7][8]}{[14]}\cdot\mathfrak{n}    \\
6.& \mathcal{H}^{11a19}_{[2,1]} - \mathcal{H}^{11a25}_{[2,1]} &=& A^{-7}\cdot f(A,q)\cdot\dfrac{[14]}{[2][7]}\cdot\mathfrak{n}   \\
7.& \mathcal{H}^{11a24}_{[2,1]} - \mathcal{H}^{11a26}_{[2,1]} &=& A^{-1} \cdot f(A,q)\cdot\dfrac{[14]}{[2][7]}\cdot\mathfrak{n}  \\
8.& \mathcal{H}^{11a251}_{[2,1]} - \mathcal{H}^{11a253}_{[2,1]} &=& A^{-1} \cdot f(A,q)\cdot\dfrac{[14]}{[2][7]}\cdot\mathfrak{n}  \\
9.& \mathcal{H}^{11a252}_{[2,1]} - \mathcal{H}^{11a254}_{[2,1]} &=& A^{-5} \cdot f(A,q) \cdot\dfrac{[14]}{[2][7]}\cdot\mathfrak{n} \\
10.& \mathcal{H}^{11n34}_{[2,1]} - \mathcal{H}^{11n42}_{[2,1]} &=& A^{3} \cdot f(A,q)\cdot\dfrac{[14]}{[2][7]}\cdot\mathfrak{n}   \\
11.& \mathcal{H}^{11n35}_{[2,1]} - \mathcal{H}^{11n43}_{[2,1]} &=& A^{19} \cdot f(A,q)\cdot\mathfrak{n}  \\
12.& \mathcal{H}^{11n36}_{[2,1]} - \mathcal{H}^{11n44}_{[2,1]} &=& A^{-9}\cdot f(A,q)\cdot\mathfrak{n}   \\
13.& \mathcal{H}^{11n39}_{[2,1]} - \mathcal{H}^{11n45}_{[2,1]} &=& A^{-3}\cdot f(A,q)\cdot\dfrac{[14]}{[2][7]}\cdot\mathfrak{n} \\
14.& \mathcal{H}^{11n40}_{[2,1]} - \mathcal{H}^{11n46}_{[2,1]} &=& A^{13} \cdot f(A,q)\cdot\mathfrak{n}  \\
15.& \mathcal{H}^{11n41}_{[2,1]} - \mathcal{H}^{11n47}_{[2,1]} &=& A^{-15}\cdot f(A,q)\cdot\mathfrak{n}  \\
16.& \mathcal{H}^{11n151}_{[2,1]} - \mathcal{H}^{11n152}_{[2,1]}  &=& A^{-9} \cdot f(A,q)\cdot\dfrac{[14]}{[2][7]}\cdot\mathfrak{n}
\end{array}
\end{equation}
where, for the sake of brevity, we introduced a standard factor $\mathfrak{n}:=\dfrac{[3]^2[14]}{[2][7]}$. Among these pairs the first 5 are pretzel knots.

 The appearance of the general factor (\ref{ffactor}) can be explained with use of the properties of quantum groups. Since $A=q^N$ for the quantum group $U_q(sl_N)$,  $D_{-i}$ is equal to zero for the quantum group $U_q(sl_i)$. This leads to the obvious conjecture that the difference should include $D_0$, $D_{-1}$ and $D_{-2}$, because mutants are not distinguished by the corresponding polynomials: multiplicities do not exist in this case. It happens so that the difference also do not exist for the $sl_3$. The replace $q\rightarrow -q^{-1}$ in the HOMFLY-PT polynomial corresponds to the transposition of the Young diagram associated with the representation. Since $[2,1]$ is self-transposed, the difference also includes $D_{1}$, $D_{2}$ and $D_{3}$. A similar structure should also exist for other representations, which we observe in the next sections.

\section{Mutant knots in representation $[3,1]$}

In representation $[3,1]$, the exclusive Racah matrices are unknown, and the braid representations of mutant knots are at least 4-strand, hence, the knot invariants are difficult to evaluate. This is why we used the classical RT approach described in the section \ref{CRT} to obtain the knot polynomials. However, this approach allows us to evaluate the knot invariants only at particular values of $N$ of the group $U_q(sl_N)$. Interestingly, we have managed to calculate the differences between HOMFLY-PT polynomials for some pairs of mutant knots for many different values of $N$ and extrapolated the answers to arbitrary $N$. We also used the connection between the HOMFLY-PT polynomials in representations corresponding to transposed Young diagrams to infer an analog of eqns.(\ref{gen21}), (\ref{ffactor}) for representation $[3,1]$. Note that the transpose of $[3,1]$ is $[2,1,1]$, and the colored HOMFLY-PT polynomials obey
\begin{equation}
H_{[3,1]}(A,q) = H_{[2,1,1]}(A,-q^{-1}).
\end{equation}
A counterpart of (\ref{gen21}), (\ref{ffactor}) in the case of $[3,1]$ is
\begin{equation}
\Delta \mathcal{H}_{[3,1]}^{mutant} = M(A,q) \cdot D_4 D_3D_2D_1D_0^2D_{-1}D_{-2}.
\end{equation}
The dependence on $N$ in this case is more complicated. Here $M$ depends on two variables $A$ and $q$ whereas, in the  $[2,1]$ case, we observed all the  $N$-dependence contained in  $f(A,q)$ and the factor  $A^{\gamma}$. The product of differentials can be partly explained by the properties of quantum groups. As in the case of $[2,1]$, we get $D_0$, $D_{-1}$ and $D_{-2}$ for representation $[3,1]$ and $D_0$, $D_{1}$ and $D_{2}$ for the transposed representation $[2,1,1]$. In the case of $sl_3$, representation $[2,1,1]$ coincides with $[1]$, which means vanishing the difference and emerging the differential $D_3$.

 For the pretzel mutant knots, we have managed to evaluate the invariants in representation $[3,1]$ of $U_q(sl_N)$ for  the values of $N=3,4,5,6,7$ and for representation $[2,1,1]$ of $U_q(sl_N)$ for $N=5,6,7,8$. Non-pretzel mutant knots are harder to deal with, thus we have managed to evaluate the invariants only in representation $[3,1]$ of $U_q(sl_3)$ and $U_q(sl_4)$ groups.

 We have been able to present the extrapolation because the degrees of the polynomials in the variable $A^2$ are smaller then the number of known points. We also have used the information about differentials for different ranks of groups. In the table below, we demonstrate how the differentials were extracted from the differences. In the column ``factors'', we listed the largest common divisor of differences of the HOMFLY-PT polynomials of different knots for certain values of $N$. The answers in the table are reduced invariants, i.e. they are normalized by the unknot, $\mathcal{H}_{[3,1]}^{{\rm unknot}} = \frac{D_2D_1D_0D_{-1}}{[4][2]}$.
	\begin{center}
		\begin{tabular}{|c|c|c|c|c|}
            \hline
			& \multicolumn{2}{c|}{ } & \multicolumn{2}{c|}{ } \\
			& \multicolumn{2}{c|}{[3,1]} & \multicolumn{2}{c|}{[2,1,1]} \\
			& \multicolumn{2}{c|}{ } & \multicolumn{2}{c|}{ } \\
			\hline
			&&&&\\
			group  & factors & differentials & factors & differentials \\
			\hline
			&&&&\\
			$sl_3$ & $[9][8]^2[7][6][4]^3[2]$ & $[7][6][3][1]$ & &\\
			&&&&\\
			\hline
			&&&&\\
			$sl_4$ & $[8][7][4]^3[2]^3$ & $[8][7][4][2]$ & &\\
			&&&&\\
			\hline
			&&&&\\
			$sl_5$ & $[9][8][6][5][4]^2[3][2]^3$ & $[9][8][5][3]$ & $[8]^2[7][5][4]^3[2]^2 $ & $[7][5][2][1]$ \\
			&&&&\\
			\hline
			&&&&\\
			$sl_6$ & $[10][9][6][5][4]^3[2]^2$ & $[10][9][6][4]$ & $[9][8][6][4]^2[2]^3$ & $[8][6][3][2]$ \\
			&&&&\\
			\hline
			&&&&\\
			$sl_7$ &$[11][10][7][5][4]^5[2]$ &$[11][10][7][5]$ & $[9][7][6][4]^3[3][2]^3$ & $[9][7][4][3]$\\
			&&&&\\
			\hline
				&&&&\\
			$sl_8$ & & & $[10][8][5][4]^3[2]^2$ & $[10][8][5][4]$\\
			&&&&\\
			\hline
			& \multicolumn{2}{c|}{} & \multicolumn{2}{c|}{} \\
			result & \multicolumn{2}{c|}{$D_{4}D_{3}D_0 D_{-2}$} & \multicolumn{2}{c|}{$D_{2}D_{0}D_{-3} D_{-4}$} \\
			& \multicolumn{2}{c|}{} & \multicolumn{2}{c|}{} \\
			\hline
		\end{tabular}
	\end{center}
 We also made an additional check of the extrapolated results using the genus\footnote{One has not confuse the genus of the pretzel knot with this genus expansion, which is an expansion at the vicinity of point $q=1$ in powers of $z = q-q^{-1}$.} expansion of the HOMFLY-PT polynomials.

\subsection{Genus expansion}

The HOMFLY-PT polynomial $H$ is a polynomial in two variables $A$ and $q$. One can expand it in powers of $z = q-q^{-1}$. This procedure is called genus expansion \cite{GE1,GE2,GE3}:
\begin{eqnarray}
H_Q^{\K}(A,q) = {_{_0}\bar{\sp}}_{_Q}^{\K}(A) + {_{_1}}\bar{\sp}_{_Q}^{\K}(A)\cdot z +  {_{_2}}\bar{\sp}_{_Q}^{\K}(A)\cdot z^2 +  {_{_3}}\bar{\sp}_{_Q}^{\K}(A)\cdot z^3 + {_{_4}}\bar{\sp}_{_Q}^{\K}(A)\cdot z^4 + ...
\end{eqnarray}
As a result, we split the dependence on the representation $Q$ and on the knot $\K$ in each order of this expansion. The dependence on the knot $\K$ is now contained in $\sigma^{\K}$, and $\varphi_{_Q}(T)$ depend on the representation $Q$ only:
\begin{eqnarray}
{_{_0}}\bar{\sigma}^{\K}_{_Q} &{=}& \Big(\sigma^{\K}_{_{[1]}}\Big)^{|Q|} \nn \\
{_{_1}}\bar{\sigma}^{\K}_{_Q} &=& \Big(\sigma^{\K}_{_{[1]}}\Big)^{|Q|-2}{\cdot}
{_{_1}}\sigma^{\K}_{_{[2]}}{\cdot}\varphi_{_Q}([2]) \nn \\
{_{_2}}\bar{\sigma}^{\K}_{_Q} &=& \Big(\sigma^{\K}_{_{[1]}}\Big)^{|Q|-4}{\cdot}
\Big( {_{_{2}}}\sigma^{\K}_{_{[1]}}\varphi_{_Q}([1]) +
{_{_2}}\sigma^{\K}_{_{[11]}}\varphi_{_Q}([11]) + {_{_2}}\sigma^{\K}_{_{[3]}}\varphi_{_Q}([3])
+ {_{_2}}\sigma^{\K}_{_{[22]}}\varphi_{_Q}([22]) \Big) \nn \\
{_{_3}}\bar{\sigma}^{\K}_{_Q} &=& \Big(\sigma^{\K}_{_{[1]}}\Big)^{|Q|-6}{\cdot}
\Big( {_{_3}}\sigma^{\K}_{_{[2]}}\varphi_{_Q}([2]) +
{_{_3}}\sigma^{\K}_{_{[21]}}\varphi_{_Q}([21]) + {_{_3}}\sigma^{\K}_{_{[4]}}\varphi_{_Q}([4])
+ {_{_3}}\sigma^{\K}_{_{[211]}}\varphi_{_Q}([211]) + {_{_3}}\sigma^{\K}_{_{[32]}}\varphi_{_Q}([32])
+ {_{_3}}\sigma^{\K}_{_{[222]}}\varphi_{_Q}([222]) \Big) \nn \\
{_{_4}}\bar{\sigma}^{\K}_{_Q} &=& \Big(\sigma^{\K}_{_{[1]}}\Big)^{|Q|-8}{\cdot}
\Big( {_{_4}}\sigma^{\K}_{_{[1]}}\varphi_{_Q}([1]) + {_{_4}}\sigma^{\K}_{_{[11]}}\varphi_{_Q}([11])
+ {_{_4}}\sigma^{\K}_{_{[3]}}\varphi_{_Q}([3]) + {_{_4}}\sigma^{\K}_{_{[111]}}\varphi_{_Q}([111])
+ \nonumber \\
&+& {_{_4}}\sigma^{\K}_{_{[31]}}\varphi_{_Q}([31]) + {_{_4}}\sigma^{\K}_{_{[22]}}\varphi_{_Q}([22])
+ {_{_4}}\sigma^{\K}_{_{[1111]}}\varphi_{_Q}([1111]) + {_{_4}}\sigma^{\K}_{_{[5]}}\varphi_{_Q}([5])
+ {_{_4}}\sigma^{\K}_{_{[311]}}\varphi_{_Q}([311]) + {_{_4}}\sigma^{\K}_{_{[221]}}\varphi_{_Q}([221])
+ \nonumber \\
&+& {_{_4}}\sigma^{\K}_{_{[42]}}\varphi_{_Q}([42])
+ {_{_4}}\sigma^{\K}_{_{[33]}}\varphi_{_Q}([33]) + {_{_4}}\sigma^{\K}_{_{[2211]}}\varphi_{_Q}([2211])
+ {_{_4}}\sigma^{\K}_{_{[322]}}\varphi_{_Q}([322])
+  {_{_4}}\sigma^{\K}_{_{[2222]}}\varphi_{_Q}([2222]) \Big)
\label{spe1}
\end{eqnarray}
The numerical coefficients $\varphi_Q(T)$ are proportional to the characters of symmetric groups, see \cite{MMN1,MMN2} for details.

Using results for the HOMFLY-PT polynomials in representations $[1]$, $[2]$, $[3]$, $[2,1]$, $[4]$ and $[2,2]$ for the pretzel mutant knots, we found all $\sigma$'s that are included into the expansion of $[3,1]$ representation. It allowed us to make an additional check of the polynomials that we got in representation $[3,1]$ of $U_q(sl_N)$.	

\subsection{11-crossing pretzel mutant knots \label{4447}}
The answers are very cumbersome, thus we provide only one example here, the remaining being listed in Appendix \ref{11cr_app} and on the site \cite{knotebook}.

The normalized difference of $[3,1]$ colored HOMFLY-PT invariants for mutant pair $11a44$ and $11a47$ is
\begin{dmath*}
	\Delta H_{[3,1]}^{11a44-11a47} = \{q\}^8 \,[4]^2\,[2]\,D_4 \,D_3\,D_0\,D_{-2}\,A^{-4}q^{-42}\left(A^8 q^{86}-A^8 q^{84}-2 A^6 q^{84}+A^8 q^{82}+3 A^6 q^{82}+A^4 q^{82}-2 A^{10} q^{80}-A^8 q^{80}-3 A^6 q^{80}-2 A^4 q^{80}+2 A^{10} q^{78}+5 A^8 q^{78}+2 A^4 q^{78}-5
	A^8 q^{76}-4 A^6 q^{76}+A^4 q^{76}+A^{12} q^{74}-2 A^{10} q^{74}+A^8 q^{74}+4 A^6 q^{74}+3 A^4 q^{74}-A^{12} q^{72}-A^{10} q^{72}+10 A^8 q^{72}-4 A^6 q^{72}-6 A^4
	q^{72}-2 A^2 q^{72}-A^{12} q^{70}+A^{10} q^{70}-2 A^8 q^{70}-5 A^6 q^{70}+11 A^4 q^{70}+5 A^2 q^{70}+3 A^{12} q^{68}-6 A^8 q^{68}-10 A^6 q^{68}-4 A^4 q^{68}-7 A^2
	q^{68}-A^{12} q^{66}-5 A^{10} q^{66}+15 A^8 q^{66}+20 A^6 q^{66}+8 A^4 q^{66}+A^2 q^{66}-2 A^{12} q^{64}-4 A^{10} q^{64}-6 A^8 q^{64}-25 A^6 q^{64}-A^4 q^{64}+2
	A^2 q^{64}+5 A^{12} q^{62}+14 A^{10} q^{62}+16 A^8 q^{62}-15 A^6 q^{62}+A^4 q^{62}-8 A^2 q^{62}+q^{62}-4 A^{12} q^{60}-18 A^{10} q^{60}-8 A^8 q^{60}+19 A^6
	q^{60}+24 A^4 q^{60}+2 A^2 q^{60}-3 q^{60}-A^{10} q^{58}+3 A^8 q^{58}-27 A^6 q^{58}-9 A^4 q^{58}+5 q^{58}+6 A^{12} q^{56}+13 A^{10} q^{56}+36 A^8 q^{56}-9 A^6
	q^{56}-4 A^4 q^{56}-14 A^2 q^{56}-2 q^{56}-6 A^{12} q^{54}-24 A^{10} q^{54}-20 A^8 q^{54}+10 A^6 q^{54}+39 A^4 q^{54}+8 A^2 q^{54}-4 q^{54}+2 A^{12} q^{52}+6
	A^{10} q^{52}-6 A^8 q^{52}-58 A^6 q^{52}-13 A^4 q^{52}+A^2 q^{52}+15 q^{52}+6 A^{12} q^{50}+13 A^{10} q^{50}+61 A^8 q^{50}+41 A^6 q^{50}+11 A^4 q^{50}-40 A^2
	q^{50}-19 q^{50}-9 A^{12} q^{48}-37 A^{10} q^{48}-44 A^8 q^{48}-23 A^6 q^{48}+47 A^4 q^{48}+46 A^2 q^{48}+15 q^{48}+8 A^{12} q^{46}+24 A^{10} q^{46}+28 A^8
	q^{46}-88 A^6 q^{46}-64 A^4 q^{46}-38 A^2 q^{46}+5 q^{46}-A^{12} q^{44}-7 A^{10} q^{44}+46 A^8 q^{44}+97 A^6 q^{44}+100 A^4 q^{44}-9 A^2 q^{44}-24 q^{44}-3 A^{12}
	q^{42}-23 A^{10} q^{42}-62 A^8 q^{42}-89 A^6 q^{42}-22 A^4 q^{42}+32 A^2 q^{42}+35 q^{42}+6 A^{12} q^{40}+22 A^{10} q^{40}+72 A^8 q^{40}-10 A^6 q^{40}-40 A^4
	q^{40}-63 A^2 q^{40}-18 q^{40}-4 A^{12} q^{38}-19 A^{10} q^{38}-4 A^8 q^{38}+30 A^6 q^{38}+107 A^4 q^{38}+37 A^2 q^{38}-7 q^{38}+2 A^{12} q^{36}-20 A^8 q^{36}-76
	A^6 q^{36}-38 A^4 q^{36}-8 A^2 q^{36}+29 q^{36}+A^{12} q^{34}+A^{10} q^{34}+45 A^8 q^{34}+22 A^6 q^{34}+A^4 q^{34}-46 A^2 q^{34}-26 q^{34}-A^{12} q^{32}-7 A^{10}
	q^{32}-5 A^8 q^{32}-3 A^6 q^{32}+54 A^4 q^{32}+40 A^2 q^{32}+8 q^{32}+A^{12} q^{30}-A^{10} q^{30}-4 A^8 q^{30}-52 A^6 q^{30}-26 A^4 q^{30}-18 A^2 q^{30}+17
	q^{30}+A^{10} q^{28}+26 A^8 q^{28}+18 A^6 q^{28}+21 A^4 q^{28}-24 A^2 q^{28}-21 q^{28}-4 A^{10} q^{26}-10 A^8 q^{26}-9 A^6 q^{26}+30 A^4 q^{26}+21 A^2 q^{26}+10
	q^{26}-A^{10} q^{24}+8 A^8 q^{24}-26 A^6 q^{24}-26 A^4 q^{24}-23 A^2 q^{24}+9 q^{24}+A^{10} q^{22}+12 A^8 q^{22}+9 A^6 q^{22}+24 A^4 q^{22}+5 A^2 q^{22}-12
	q^{22}-2 A^{10} q^{20}-9 A^8 q^{20}-15 A^6 q^{20}+12 A^4 q^{20}+4 A^2 q^{20}+4 q^{20}+8 A^8 q^{18}-6 A^4 q^{18}-19 A^2 q^{18}+5 q^{18}+2 A^8 q^{16}-A^6 q^{16}+5
	A^4 q^{16}+9 A^2 q^{16}-3 q^{16}-A^8 q^{14}-8 A^6 q^{14}+6 A^4 q^{14}+2 A^2 q^{14}+A^8 q^{12}-A^6 q^{12}+2 A^4 q^{12}-10 A^2 q^{12}+q^{12}+A^8 q^{10}+3 A^6
	q^{10}+A^4 q^{10}+3 A^2 q^{10}-5 A^6 q^8-2 A^4 q^8+A^2 q^8+4 A^4 q^6-2 A^2 q^6+A^6 q^4-A^6 q^2-A^4 q^2+A^4\right)
\end{dmath*}

The following coefficients of the genus expansion do not distinguish between mutant knots, and for knots $11a44$ and $11a47$ are equal to
\begin{equation}
\begin{array}{lll}
\sigma_{_{[1]}}& =&-A^{-4} (5 A^6-13 A^4+9 A^2-2) ,  \\ &&\\
_{_1}\sigma_{_{[2]}}& =& A^{-8} [\left(A^2-1\right) \left(53 A^{10}-116 A^8+29 A^6+70 A^4-53 A^2+11\right)], \\ &&\\
_{_2}\sigma_{_{[1]}}&=&-\sigma_{_{[1]}}^3 (7 A^6-22 A^4+15 A^2-3)A^{-4}, \\&&\\
 _{_2}\sigma_{_{[1,1]}}&=&\sigma_{_{[1]}}^2 [299 A^{12}-1293 A^{10}+2303 A^8-2159 A^6+1187 A^4-382 A^2+57](2 A^8)^{-1}, \\&&\\
 _{_2}\sigma_{_{[3]}}&=& -\sigma_{_{[1]}} [\left(A^2-1\right)^2 \left(2011 A^{14}-5594 A^{12}+5010 A^{10}-2501 A^8+2509 A^6-2497 A^4+1139 A^2-188\right)](2 A^{12})^{-1}, \\&&\\
  _{_2}\sigma_{_{[2,2]}}&=& _{_1}\sigma_{_{[2]}}^2,\\
  &&\\
  _{_3} \sigma_{_{[2]}} & = & \sigma_{_{[1]}}^4 [\left(A^2-1\right) \left(3125 A^{10}-6708 A^8+197 A^6+5998 A^4-3989 A^2+811\right)](8 A^8)^{-1},\\&&\\
   _{_3} \sigma_{_{[2,1]}} & = &- \sigma_{_{[1]}}^3 [\left(A^2-1\right) \left(13936 A^{16}-56799 A^{14}+82137 A^{12}
   -34516 A^{10}-35635 A^8+55190 A^6\right. \\&~&\left. -32764 A^4+10033 A^2-1318\right)](3 A^{12})^{-1},\\&&\\
    _{_3} \sigma_{_{[4]}} & = & \sigma_{_{[1]}}^2 L(A^2-1)(6 A^{16})^{-1}\\&&\\
     _{_3} \sigma_{_{[2,1,1]}} & =& (_{_1}\sigma_{_{[2]}}) \cdotp (_{_2}\sigma_{_{[1,1]}})  ,\\
     &&\\
     _{_4}\sigma_{_{[1]}} & = &- \sigma_{_{[1]}}^7 \left(2 A^2-1\right) \left(2 A^4-7 A^2+1\right)A^{-4} ,\\&&\\
      _{_4}\sigma_{_{[1,1]}} & = & \sigma_{_{[1]}}^6 [1702 A^{12}-7432 A^{10}+13357 A^8-12798 A^6+7557 A^4-2712 A^2+444] (2 A^8)^{-1},\\&&\\
       _{_4}\sigma_{_{[1,1,1,1]}} & = &3 (_{_2}\sigma_{_{[1,1]}})^2 ,\\
 \end{array}
\end{equation}
where
\begin{equation}
\begin{array}{lll}
L&=&136412 A^{24}-844179 A^{22}+2169523 A^{20}-2942698 A^{18}+2101236 A^{16}-361818 A^{14}-919414 A^{12}+\\&&1347985 A^{10}-1168221 A^8+696262 A^6-269099 A^4+59776 A^2-5765.\\
\end{array}
\end{equation}
The difference between the HOMFLY polynomials of knots $11a44$ and $11a47$ emerges only in the forth order of genus expansion:
\begin{equation}
\begin{array}{lll}
_{_4}\sigma_{_{[3]}}^{11a44}& =& - \sigma_{_{[1]}}^5 [\left(A^2-1\right)^2 \left(126502 A^{14}-330767 A^{12}+272505 A^{10}-161495 A^8+212800 A^6-203005 A^4+89447 A^2-15038\right)]({6 A^{12}})^{-1},\\&&\\
_{_4}\sigma_{_{[1,1,1]}}^{11a44} & = & \sigma_{_{[1]}}^5[-115078 A^{18}+681937 A^{16}-1808204 A^{14}+2851057 A^{12}-2997653 A^{10}+2224199 A^8-1177844 A^6+430039 A^4-98241 A^2+10706] ({6 A^{12}})^{-1} ,\\&&\\
_{_4} \sigma_{_{[3,1]}}^{11a44} & =&\sigma_{_{[1]}}^4 M_1({8 A^{16}})^{-1},\\&&\\
_{_4}\sigma_{_{[2,2]}}^{11a44}&=&\sigma_{_{[1]}}^4 M_2 ({12 A^{16}})^{-1}	\\
&&\\
_{_4}\sigma_{_{[3]}}^{11a47}& =& _{_4}\sigma_{_{[3]}}^{11a44}-\sigma_{_{[1]}}^5[12 \left(A^2-1\right)^7]{A^{-8}}\\&&\\
_{_4}\sigma_{_{[1,1,1]}}^{11a47} & = &_{_4}\sigma^{11a44}_{_{[1,1,1]}}+\sigma_{_{[1]}}^5[24 \left(A^2-1\right)^7]{A^{-8}}\\&&\\
_{_4}\sigma_{_{[3,1]}}^{11a47} & = &_{_4}\sigma_{_{[3,1]}}^{11a44} + \sigma_{_{[1]}}^5 [12 \left(A^2-1\right)^7]{A^{-8}}\\&&\\
_{_4}\sigma_{_{[2,2]}}^{11a47}&=& _{_4}\sigma_{_{[2,2]}}^{11a44}-\sigma_{_{[1]}}^5 32 \left(A^2-1\right)^7{A^{-8}},	\\
\end{array}
\end{equation}
where
\begin{equation}
\begin{array}{lll}
M_1 &=&1119235 A^{24}-7225039 A^{22}+19907264 A^{20}-30800410 A^{18}+30670905 A^{16}-23997490 A^{14}+\\
&& +20086857 A^{12}-18271527 A^{10}+13724564 A^8-7265762 A^6+2522613
A^4-519606 A^2+48372, \\ \\
M_2 &=& 2219023 A^{24}-13508232 A^{22}+32990983 A^{20}-38826012 A^{18}+17251727 A^{16}+6217024 A^{14}-\\
&&-3560525 A^{12}-13683778 A^{10}+20342527 A^8-13652158 A^6+5218892
A^4-1113552 A^2+104153.\\
\end{array}
\end{equation}

\section{Mutant knots in representation $[4,2]$}

In \cite{Mor2}, the differences for two pairs of knots in representation $[4,2]$ of $U_q(sl_3)$ were calculated. We have reproduced these results using the classical RT approach \ref{CRT}. Further, we calculated these mutant knot pair differences for the group $U_q(sl_4)$. We have also performed this for many other mutant pairs.

\subsection{Morton mutant pairs}

For two pairs of the pretzel knots mentioned in the Morton paper \cite{Mor2}, $K(3,3,3,-3,-3)-K(3,3,-3,3,-3)$ and $K(1,3,3,-3,-3)-K(1,3,-3,3,-3)$, the differences for the HOMFLY-PT polynomials in representation $[4,2]$ of the group $U_q(sl_3)$ are  as follows (obtained using the RT approach):
\begin{dmath*}
	\Delta \mathcal{H}_{[4,2]}^{K(1,3,3,-3,-3)-K(1,3,-3,3,-3)} =2 q^{-136} \left(q^2-1\right)^{18} \left(q^3+q\right)^{10} \left(q^4+1\right)^4 \left(q^4-q^2+1\right)^5 \left(q^4+q^2+1\right)^7 \left(q^8+q^6+q^4+q^2+1\right)^3
	\left(q^{20}-q^{16}+q^{14}+q^{12}-q^{10}+q^8+q^6-q^4+1\right)^2 \left(q^{20}+q^{18}+q^{16}+q^{14}+2 q^{12}+2 q^{10}+2 q^8+q^6+q^4+q^2+1\right)
\end{dmath*}
\begin{dmath*}
	\Delta \mathcal{H}_{[4,2]}^{K(3,3,3,-3,-3)-K(3,3,-3,3,-3)}=2 q^{-174} \left(q^2-1\right)^{18} \left(q^2+1\right)^{10} \left(q^4+1\right)^4 \left(q^4-q^2+1\right)^5 \left(q^4+q^2+1\right)^7 \left(q^8+q^6+q^4+q^2+1\right)^3
	\left(q^{20}-q^{16}+q^{14}+q^{12}-q^{10}+q^8+q^6-q^4+1\right)^2 \left(2 q^{84}-2 q^{82}-2 q^{80}+5 q^{78}-5 q^{76}-2 q^{74}+14 q^{72}-14 q^{70}-12 q^{68}+35
	q^{66}-13 q^{64}-33 q^{62}+46 q^{60}+4 q^{58}-54 q^{56}+34 q^{54}+24 q^{52}-49 q^{50}+15 q^{48}+18 q^{46}-37 q^{44}+19 q^{42}+15 q^{40}-39 q^{38}+23 q^{36}+23
	q^{34}-40 q^{32}+12 q^{30}+37 q^{28}-26 q^{26}-11 q^{24}+30 q^{22}+q^{20}-16 q^{18}+10 q^{16}+12 q^{14}-5 q^{12}-3 q^{10}+8 q^8+3 q^6-2 q^4+q^2+2\right)
	\end{dmath*}	
These differences coincide with the result in \cite{Mor2} after substituting $q \rightarrow q^{-2}$ and multiplying by the unknot HOMFLY-PT polynomial ($\mathcal{H}_{[4,2]}^{\text{unknot}} = \frac{[6][3]}{[2]}$).
We also managed to evaluate the $[4,2]$ colored HOMFLY-PT polynomial for these four knots, and
the results can be found in Appendix \ref{MorKnots}.

For the group $U_q(sl_4)$ with unknot invariant $$\mathcal{H}_{[4,2]}^{\text{unknot}} = \frac{[7][6][4][3]}{[2]^2}$$,
 the non-normalized differences are
\begin{dmath*}
	\Delta \mathcal{H}_{[4,2]}^{K(1,3,3,-3,-3)-K(1,3,-3,3,-3)} = 2 q^{-156} (q-1)^{18} (q+1)^{18} \left(q^2+1\right)^{10} \left(q^2-q+1\right)^5 \left(q^2+q+1\right)^5 \left(q^4+1\right)^4 \left(q^4-q^2+1\right)^2
	\left(q^4-q^3+q^2-q+1\right)^3 \left(q^4+q^3+q^2+q+1\right)^3 \left(q^6-q^3+1\right) \left(q^6+q^3+1\right) \left(q^6-q^5+q^4-q^3+q^2-q+1\right)
	\left(q^6+q^5+q^4+q^3+q^2+q+1\right) \left(q^8+1\right) \left(q^8-q^6+q^4-q^2+1\right)^2 \left(q^{40}+q^{36}+2 q^{34}+q^{32}+q^{30}+3 q^{28}+q^{26}+2 q^{24}+3
	q^{22}+2 q^{20}+3 q^{18}+2 q^{16}+q^{14}+3 q^{12}+q^{10}+q^8+2 q^6+q^4+1\right)^2
\end{dmath*}
\begin{dmath*}
	\Delta \mathcal{H}_{[4,2]}^{K(3,3,3,-3,-3)-K(3,3,-3,3,-3)}= 2 q^{-224} (q-1)^{18} (q+1)^{18} \left(q^2+1\right)^{10} \left(q^2-q+1\right)^5 \left(q^2+q+1\right)^5
	\left(q^4+1\right)^4 \left(q^4-q^2+1\right)^2 \left(q^4-q^3+q^2-q+1\right)^3
	\left(q^4+q^3+q^2+q+1\right)^3 \left(q^6-q^3+1\right) \left(q^6+q^3+1\right)
	\left(q^6-q^5+q^4-q^3+q^2-q+1\right) \left(q^6+q^5+q^4+q^3+q^2+q+1\right) \left(q^8+1\right)
	\left(q^8-q^6+q^4-q^2+1\right)^2 \left(q^{40}+q^{36}+2 q^{34}+q^{32}+q^{30}+3 q^{28}+q^{26}+2
	q^{24}+3 q^{22}+2 q^{20}+3 q^{18}+2 q^{16}+q^{14}+3 q^{12}+q^{10}+q^8+2 q^6+q^4+1\right)^2
	\left(2 q^{84}-4 q^{82}-2 q^{80}+10 q^{78}-5 q^{76}-10 q^{74}+17 q^{72}-q^{70}-25 q^{68}+22
	q^{66}+9 q^{64}-21 q^{62}+7 q^{60}+3 q^{58}+7 q^{56}-22 q^{54}-q^{52}+46 q^{50}-43 q^{48}-7
	q^{46}+43 q^{44}-41 q^{42}+11 q^{40}+17 q^{38}-24 q^{36}+23 q^{34}-12 q^{32}-9 q^{30}+20
	q^{28}-17 q^{26}+7 q^{24}+7 q^{22}-12 q^{20}+10 q^{18}-4 q^{16}-6 q^{14}+9 q^{12}-2 q^{10}-2
	q^8+4 q^6-2 q^4-2 q^2+2\right)
\end{dmath*}	

\subsection{11-crossing mutant knots}

We have evaluated the answers for $11$-crossing mutant knots for the both $U_q(sl_3)$ and $U_q(sl_4)$ groups. For  the $U_q(sl_3)$ group, the differences have  a  nice form similar to (\ref{mutlist}):
\begin{equation}
\begin{array}{llcl}
1. &	\Delta H_{[4,2]}^{11a19-11a25}   & = & -2\{q\}^{14}\, \dfrac{[9]^3}{[3]^3} \, [8]\, [7]\,[5]^2\,[4]^3\,[2]^4\,q^{-60}\,\left(q^8-q^4+1\right)^2 \left(q^{12}+q^8-1\right)^2,\\
2. &	\Delta H_{[4,2]}^{11a24-11a26}   & = & -2\{q\}^{14}\, \dfrac{[9]^3}{[3]^3} \, [8]\, [7]\,[5]^2\,[4]^3\,[2]^4\,q^{-28}\,\left(q^8-q^4+1\right)^2 \left(q^{12}-q^4-1\right) \left(q^{12}+q^8-1\right),\\
3. &	\Delta H_{[4,2]}^{11a44-11a47}   & = &  2\{q\}^{14}\, [12] \,\dfrac{[9]^2}{[3]^2} \, [8]\,[7]\,[5]^2\,[4]\,[2]^4\,q^{-6}\left(q^{12}+q^8-1\right),\\
4. & 	\Delta H_{[4,2]}^{11a57-11a231}  & = & -2\{q\}^{14}\, [12] \,\dfrac{[9]^2}{[3]^2} \, [8]\,[7]\,[5]^2\,[4]\,[2]^4\,q^{26}\,\left(q^{12}-q^4-1\right),\\
5. &	\Delta H_{[4,2]}^{11a251-11a253} & = & -2\{q\}^{14}\, \dfrac{[9]^3}{[3]^3} \, [8]\, [7]\,[5]^2\,[4]^3\,[2]^4\,q^{-12}\,\left(q^8-q^4+1\right)^2 \left(q^{12}-q^4-1\right) \left(q^{12}+q^8-1\right),\\
6. &	\Delta H_{[4,2]}^{11a252-11a254} & = & -2\{q\}^{14}\, \dfrac{[9]^3}{[3]^3} \, [8]\, [7]\,[5]^2\,[4]^3\,[2]^4\,q^{-44}\,\left(q^8-q^4+1\right)^2 \left(q^{12}+q^8-1\right)^2,\\
7. &	\Delta H_{[4,2]}^{11n34-11n42}   & = & -2\{q\}^{14}\, \dfrac{[9]^3}{[3]^3} \, [8]\, [7]\,[5]^2\,[4]^3\,[2]^4\,q^{4}\,\left(q^8-q^4+1\right)^2 \left(q^{12}-q^4-1\right) \left(q^{12}+q^8-1\right),\\
8. &	\Delta H_{[4,2]}^{11n35-11n43}   & = & -2\{q\}^{14}\, \dfrac{[9]^3}{[3]^3} \, [8]\, [7]\,[5]^2\,[4]^3\,[2]^4\,q^{-132}\,\left(q^8-q^4+1\right)^2 \left(q^{12}+q^8-1\right)^2,\\
9. &	\Delta H_{[4,2]}^{11n36-11n44}   & = & -2\{q\}^{14}\, \dfrac{[9]^3}{[3]^3} \, [8]\, [7]\,[5]^2\,[4]^3\,[2]^4\,q^{-68}\,\left(q^8-q^4+1\right)^2 \left(q^{12}-q^4-1\right) \left(q^{12}+q^8-1\right),\\
10. &	\Delta H_{[4,2]}^{11n39-11n45}   & = & -2\{q\}^{14}\, \dfrac{[9]^3}{[3]^3} \, [8]\, [7]\,[5]^2\,[4]^3\,[2]^4\,q^{-28}\,\left(q^8-q^4+1\right)^2 \left(q^{12}+q^8-1\right)^2,\\
11. &	\Delta H_{[4,2]}^{11n40-11n46}   & = & -2\{q\}^{14}\, \dfrac{[9]^3}{[3]^3} \, [8]\, [7]\,[5]^2\,[4]^3\,[2]^4 \,q^{-100}\,\left(q^8-q^4+1\right)^2 \left(q^{12}-q^4-1\right) \left(q^{12}+q^8-1\right),\\		
12. &	\Delta H_{[4,2]}^{11n41-11n47}   & = & -2\{q\}^{14}\, \dfrac{[9]^3}{[3]^3} \, [8]\, [7]\,[5]^2\,[4]^3\,[2]^4 \,q^{-100}\,\left(q^8-q^4+1\right)^2 \left(q^{12}+q^8-1\right)^2,\\
13. &	\Delta H_{[4,2]}^{11n71-11n75}   & = & -2\{q\}^{14}\, [12] \,\dfrac{[9]^2}{[3]^2} \, [8]\,[7]\,[5]^2\,[4]\,[2]^4 \,q^{-78}\left(q^{12}+q^8-1\right),\\
14. &	\Delta H_{[4,2]}^{11n73-11n74}   & = &  2\{q\}^{14}\, [12] \,\dfrac{[9]^2}{[3]^2} \, [8]\,[7]\,[5]^2\,[4]\,[2]^4\,q^{-22}\left(q^{12}+q^8-1\right),\ \\
15. &	\Delta H_{[4,2]}^{11n76-11n78}   & = &  2\{q\}^{14}\, [12] \,\dfrac{[9]^2}{[3]^2} \, [8]\,[7]\,[5]^2\,[4]\,[2]^4\,q^{-94}\, \left(q^{12}+q^8-1\right),\\
16. &	\Delta H_{[4,2]}^{11n151-11n152} & = & -2\{q\}^{14}\, \dfrac{[9]^3}{[3]^3} \, [8]\, [7]\,[5]^2\,[4]^3\,[2]^4\, q^{-76}\left(q^8-q^4+1\right)^2 \left(q^{12}+q^8-1\right)^2.
\end{array}
\end{equation}
The differences for the $U_q(sl_4)$ group do not have such a nice form. We provide only one example here, and other mutant pairs are listed in Appendix \ref{11kn42}.
\begin{dmath}
\Delta H_{[4,2]}^{11a44-11a47} = \{q\}^{11}\,[10]\,[8]\,[5]^2\,[4]\,[2]^4\, q^{-134} \left(q^{240}+q^{238}-5 q^{236}+3 q^{234}+16 q^{232}-25 q^{230}-10 q^{228}+68 q^{226}-55 q^{224}-63 q^{222}+159 q^{220}-77 q^{218}-152 q^{216}+283 q^{214}-133 q^{212}-243
q^{210}+521 q^{208}-357 q^{206}-362 q^{204}+1071 q^{202}-861 q^{200}-577 q^{198}+2005 q^{196}-1625 q^{194}-821 q^{192}+3090 q^{190}-2510 q^{188}-924 q^{186}+4049
q^{184}-3461 q^{182}-594 q^{180}+4401 q^{178}-4299 q^{176}+747 q^{174}+3089 q^{172}-4627 q^{170}+4031 q^{168}-1571 q^{166}-3139 q^{164}+8612 q^{162}-9627
q^{160}+1978 q^{158}+10583 q^{156}-17480 q^{154}+10665 q^{152}+6090 q^{150}-20139 q^{148}+19642 q^{146}-4350 q^{144}-14450 q^{142}+22450 q^{140}-14428
q^{138}-2442 q^{136}+15829 q^{134}-18756 q^{132}+11647 q^{130}+1895 q^{128}-15789 q^{126}+21696 q^{124}-12857 q^{122}-6827 q^{120}+22591 q^{118}-21389
q^{116}+4642 q^{114}+14050 q^{112}-21105 q^{110}+13323 q^{108}+1819 q^{106}-13568 q^{104}+14750 q^{102}-6358 q^{100}-4158 q^{98}+9644 q^{96}-8167 q^{94}+2588
q^{92}+2934 q^{90}-5824 q^{88}+5141 q^{86}-1372 q^{84}-2846 q^{82}+4574 q^{80}-2799 q^{78}-832 q^{76}+3227 q^{74}-2650 q^{72}+q^{70}+2218 q^{68}-2081
q^{66}-q^{64}+1659 q^{62}-1355 q^{60}-216 q^{58}+1202 q^{56}-730 q^{54}-313 q^{52}+726 q^{50}-333 q^{48}-204 q^{46}+349 q^{44}-131 q^{42}-81 q^{40}+111 q^{38}-32
q^{36}-19 q^{34}+q^{32}+15 q^{30}+6 q^{28}-26 q^{26}+10 q^{24}+16 q^{22}-16 q^{20}-7 q^{18}+15 q^{16}-12 q^{12}+7 q^{10}+4 q^8-6 q^6+q^4+2 q^2-1\right)
\end{dmath}

\section{Conclusion}

In this paper, we have studied mutant knots using the $\mathcal{R}$-matrix in various representations. We managed to evaluate the HOMFLY-PT polynomials in non-symmetric representations $[2,1]$, $[3,1]$ and $[4,2]$ for different mutant knots. Due to the structure of $\mathcal{R}$-matrix, one can obtain the answers only for group $U_q(sl_N)$ of a concrete rank $N$ so that the results are functions of only $q$. However, in certain cases we could  infer or construct mutant knot difference for arbitrary $N$.

Particularly, using the RT approach, we obtained the differences for all 11-crossing mutant knots in representation $[3,1]$. These differences were calculated for many particular values of $N$ and then we  inferred the answer for any $N$.
 This allowed us to study the differential structure of these differences similar to the our earlier work \cite{mut21}, where representation $[2,1]$ was studied.

We also considered the Morton mutants, which are not distinguished by the representations $[2,1]$ and $[3,1]$, the lowest representation that can distinguish these mutant pairs being $[4,2]$ \cite{Mor2}. We verified the differences by explicitly computing the $[4,2]$ colored HOMFLY-PT invariants for these mutant pairs for the group  $U_q(sl_3)$.
Using the RT framework, we could compute the differences for $U_q(sl_4)$ group as well. We computed $[4,2]$ colored HOMFLY-PT polynomials for the both $U_q(sl_3)$ and $U_q(sl_4)$ groups and presented the results for all the 11-crossing mutant pairs.
Unfortunately, at the moment we were not able to evaluate these differences for higher rank $N$ of $U_q(sl_N)$ group.
 One of the ways to do this is to study the differential expansion structure of these differences in the spirit of \cite{Mila1} to get clue for a general $N$. We hope to pursue this direction in future.

\section*{Acknowledgements}
Our work is supported in part by the grant of the
Foundation for the Advancement of Theoretical Physics ``BASIS" (L.B., A.Mir, A.Mor., An.Mor., A.S.),
by  RFBR grants 19-01-00680 (A.Mir.), 19-02-00815 (A.Mor.) and 20-01-00644 (An.Mor., A.S.),
by joint grants 19-51-50008-YaF-a (L.B., A.Mir., An.Mor.), 19-51-53014-GFEN-a (L.B., A.Mir, A.Mor., An.Mor., A.S.), 18-51-05015-Arm-a (L.B., A.Mir, A.Mor., An.Mor., A.S.), 18-51-45010-IND-a (L.B., A.Mir, A.Mor., An.Mor., A.S.).
PR, SD and VKS  would like to acknowledge DST-RFBR grant (INT/RUS/RFBR/P-309) for support.
A.Mir., A.Mor. and PR also acknowledge the hospitality of KITP where some parts of this work was discussed.
The work was also partly funded by RFBR and NSFB according
to the research project 19-51-18006 (A.Mir., A.Mor., An.Mor.).

\newpage

\appendix

\section{11-crossing mutant knots \label{11cr_app}}

In this Appendix, we list our results for the differences between HOMFLY-PT polynomials of 11-crossing mutant knots and their properties for the representation $[3,1]$. These mutant pairs are divided in two groups: pretzel and ``non-pretzel'' knots. The pretzel knots due to their form are much easier to evaluate. Thus we have managed to find the polynomials for the general $U_q(sl_N)$ group, while for the ``non-pretzel'' knots we have managed to calculate only the differences in the $U_q(sl_3)$ and $U_q(sl_4)$ cases. The full list of 11-crossing mutant knots is given in section \ref{smut11}. Among those, the first five pairs are the pretzel knots.

\subsection{Pretzel mutant knots}

\paragraph{$11a44 - 11a47$}

Results are provided in section\ref{4447}.

\paragraph{$11a57 - 11a231$}

The normalized difference is

\begin{dmath*}
	\Delta H_{[3,1]}^{11a57-11a231} =\{q\}^8 \,[4]^2\,[2]\,D_4 \,D_3\,D_0\,D_{-2}\,q^{-40} \left(A^8 q^{86}-A^8 q^{84}-A^6 q^{84}+A^6 q^{82}-2 A^{10} q^{80}+3 A^8 q^{80}+2 A^{10} q^{78}-2 A^8 q^{78}-4 A^6 q^{78}+2 A^{10} q^{76}+A^8 q^{76}+3 A^6 q^{76}+A^4
	q^{76}+A^{12} q^{74}-10 A^{10} q^{74}-2 A^6 q^{74}+A^4 q^{74}-A^{12} q^{72}+8 A^{10} q^{72}+7 A^8 q^{72}-4 A^6 q^{72}-A^4 q^{72}-2 A^{12} q^{70}+3 A^{10}
	q^{70}+2 A^4 q^{70}+7 A^{12} q^{68}-16 A^{10} q^{68}-8 A^8 q^{68}-2 A^6 q^{68}+6 A^4 q^{68}-4 A^{12} q^{66}+10 A^{10} q^{66}+16 A^8 q^{66}-8 A^6 q^{66}-8 A^4
	q^{66}-2 A^2 q^{66}-8 A^{12} q^{64}-2 A^{10} q^{64}+9 A^8 q^{64}+9 A^6 q^{64}+15 A^4 q^{64}+A^2 q^{64}+18 A^{12} q^{62}-9 A^{10} q^{62}-24 A^8 q^{62}-28 A^6
	q^{62}-3 A^4 q^{62}-2 A^2 q^{62}-8 A^{12} q^{60}+15 A^{10} q^{60}+36 A^8 q^{60}+9 A^6 q^{60}-A^4 q^{60}-2 A^2 q^{60}-14 A^{12} q^{58}-34 A^{10} q^{58}-7 A^8
	q^{58}+5 A^6 q^{58}+22 A^4 q^{58}-A^2 q^{58}+30 A^{12} q^{56}+25 A^{10} q^{56}-7 A^8 q^{56}-46 A^6 q^{56}-14 A^4 q^{56}-A^2 q^{56}+q^{56}-21 A^{12} q^{54}+7
	A^{10} q^{54}+49 A^8 q^{54}+26 A^6 q^{54}+10 A^4 q^{54}-9 A^2 q^{54}-q^{54}-4 A^{12} q^{52}-56 A^{10} q^{52}-48 A^8 q^{52}-18 A^6 q^{52}+36 A^4 q^{52}+6 A^2
	q^{52}+2 q^{52}+30 A^{12} q^{50}+53 A^{10} q^{50}+35 A^8 q^{50}-44 A^6 q^{50}-36 A^4 q^{50}-10 A^2 q^{50}-33 A^{12} q^{48}-22 A^{10} q^{48}+56 A^8 q^{48}+52 A^6
	q^{48}+41 A^4 q^{48}-8 A^2 q^{48}-q^{48}+15 A^{12} q^{46}-36 A^{10} q^{46}-86 A^8 q^{46}-84 A^6 q^{46}+19 A^4 q^{46}+15 A^2 q^{46}+5 q^{46}+14 A^{12} q^{44}+55
	A^{10} q^{44}+80 A^8 q^{44}+10 A^6 q^{44}-35 A^4 q^{44}-31 A^2 q^{44}-4 q^{44}-27 A^{12} q^{42}-53 A^{10} q^{42}+13 A^8 q^{42}+36 A^6 q^{42}+73 A^4 q^{42}+12
	A^2 q^{42}+4 q^{42}+23 A^{12} q^{40}+7 A^{10} q^{40}-41 A^8 q^{40}-111 A^6 q^{40}-29 A^4 q^{40}-4 A^2 q^{40}+2 q^{40}-6 A^{12} q^{38}+21 A^{10} q^{38}+70 A^8
	q^{38}+49 A^6 q^{38}+13 A^4 q^{38}-22 A^2 q^{38}-4 q^{38}-6 A^{12} q^{36}-37 A^{10} q^{36}-20 A^8 q^{36}-16 A^6 q^{36}+47 A^4 q^{36}+14 A^2 q^{36}+6 q^{36}+12
	A^{12} q^{34}+9 A^{10} q^{34}+5 A^8 q^{34}-49 A^6 q^{34}-23 A^4 q^{34}-13 A^2 q^{34}-q^{34}-8 A^{12} q^{32}+3 A^{10} q^{32}+35 A^8 q^{32}+16 A^6 q^{32}+13 A^4
	q^{32}-14 A^2 q^{32}-q^{32}+3 A^{12} q^{30}-13 A^{10} q^{30}-9 A^8 q^{30}-19 A^6 q^{30}+41 A^4 q^{30}+11 A^2 q^{30}+5 q^{30}+2 A^{12} q^{28}+A^{10} q^{28}-A^8
	q^{28}-33 A^6 q^{28}-18 A^4 q^{28}-15 A^2 q^{28}-2 q^{28}-2 A^{12} q^{26}+3 A^{10} q^{26}+30 A^8 q^{26}+21 A^6 q^{26}+18 A^4 q^{26}-7 A^2 q^{26}+A^{12}
	q^{24}-12 A^{10} q^{24}-11 A^8 q^{24}-29 A^6 q^{24}+15 A^4 q^{24}+8 A^2 q^{24}+4 q^{24}+8 A^{10} q^{22}+12 A^8 q^{22}-13 A^6 q^{22}-8 A^4 q^{22}-12 A^2 q^{22}-2
	q^{22}-4 A^{10} q^{20}+A^8 q^{20}+10 A^6 q^{20}+22 A^4 q^{20}-A^2 q^{20}+q^{20}-4 A^{10} q^{18}+A^8 q^{18}-14 A^6 q^{18}-4 A^4 q^{18}-3 A^2 q^{18}+2 q^{18}+4
	A^{10} q^{16}+8 A^8 q^{16}-4 A^6 q^{16}+3 A^4 q^{16}-A^2 q^{16}-q^{16}-2 A^{10} q^{14}-3 A^8 q^{14}-4 A^6 q^{14}+9 A^4 q^{14}-A^2 q^{14}+2 A^8 q^{12}+2 A^4
	q^{12}-2 A^2 q^{12}+q^{12}+A^8 q^{10}-3 A^6 q^{10}-3 A^4 q^{10}-A^2 q^{10}+2 A^8 q^8+5 A^4 q^8+A^2 q^8-2 A^8 q^6-4 A^6 q^6-A^4 q^6-2 A^2 q^6+A^8 q^4+3 A^6 q^4+2
	A^4 q^4-2 A^6 q^2-A^4 q^2+A^4\right)
\end{dmath*}

The following coefficients in the genus expansion do not distinguish mutant knots, and, for knots $11a57$ and $11a231$, are
\begin{equation}
\begin{array}{lll}
\sigma_{_{[1]}}& =&-\frac{3 A^6-10 A^4+12 A^2-4}{A^2} ,  \\ &&\\
_{_1}\sigma_{_{[2]}}& =& \frac{\left(A^2-1\right) \left(25 A^{10}-79 A^8+58 A^6+66 A^4-112 A^2+40\right)}{A^4}, \\ &&\\
_{_2}\sigma_{_{[1]}}&=&-\sigma_{_{[1]}}^3 \frac{6 A^6-19 A^4+18 A^2-4}{A^2}, \\
_{_2}\sigma_{_{[1,1]}}&=&\sigma_{_{[1]}}^2 \frac{151 A^{12}-738 A^{10}+1667 A^8-2244 A^6+1862 A^4-880 A^2+186}{2 A^4}, \\
_{_2}\sigma_{_{[3]}}&=& -\sigma_{_{[1]}} \frac{\left(A^2-1\right)^2 \left(707 A^{14}-2756 A^{12}+3690 A^{10}-1942 A^8+2068 A^6-4846 A^4+4424 A^2-1304\right)}{2 A^6}, \\
_{_2}\sigma_{_{[2,2]}}&=& _{_1}\sigma_{_{[2]}}^2,\\
&&\\
_{_3}\sigma_{_{[2]}} & = & \sigma_{_{[1]}}^4 \frac{(A^2-1) (A+1) \left(2145 A^{10}-6535 A^8+4866 A^6+3722 A^4-6144 A^2+1960\right)}{8 A^4},\\
_{_3}\sigma_{_{[2,1]}} & = &- \sigma_{_{[1]}}^3 \frac{(A^2-1) L_1}{3 A^{6}},\\
_{_3}\sigma_{_{[4]}} & = & \sigma_{_{[1]}}^2 \frac{(A^2-1)L_2}{6 A^{8}}\\
_{_3}\sigma_{_{[2,1,1]}} & =& (_{_1}\sigma_{_{[2]}}) \cdotp (_{_2}\sigma_{_{[1,1]}})  ,\\
&&\\
_{_4}\sigma_{_{[1]}} & = &- \sigma_{_{[1]}}^7 \frac{4 A^6-15 A^4+10 A^2-1}{A^2},\\
_{_4}\sigma_{_{[1,1]}} & = & \sigma_{_{[1]}}^6 \frac{1311 A^{12}-5854 A^{10}+11406 A^8-13180 A^6+9671 A^4-4268 A^2+860}{2 A^4},\\
_{_4}\sigma_{_{[1,1,1,1]}} & = &3 (_{_2}\sigma_{_{[1,1]}})^2 ,\\

\end{array}
\end{equation}
where
\begin{equation}
\begin{array}{lll}
L_1&=&5628 A^{16}-28810 A^{14}+60313 A^{12}-57482 A^{10}-572 A^8+65604 A^6-75409 A^4+38764 A^2-8008 \\
L_2&=&36360 A^{22}-241248 A^{20}+663565 A^{18}-957367 A^{16}+748062 A^{14}-223682 A^{12}-356686 A^{10}+ \\&&1032510 A^8-1445080 A^6+1127144 A^4-460384 A^2+76864.\\
\end{array}
\end{equation}

The difference between HOMFLY polynomials of knots $11a44$ and $11a47$ emerges only in the forth order of genus expansion.
\begin{equation}
\begin{array}{lll}
_{_4}\sigma_{_{[3]}}^{11a57}& =& - \sigma_{_{[1]}}^5\frac{\left(A^2-1\right)^2 \left(64625 A^{14}-233582 A^{12}+298194 A^{10}-152101 A^8+121231 A^6-267661 A^4+241916 A^2-71678\right)}{6 A^6},\\
_{_4}\sigma_{_{[1,1,1]}}^{11a57} & = & \sigma_{_{[1]}}^5\frac{-51842 A^{18}+337308 A^{16}-1034565 A^{14}+2005923 A^{12}-2753070 A^{10}+2792872 A^8-2086847 A^6+1096939 A^4-362676 A^2+56372}{6 A^6} ,\\
_{_4}\sigma_{_{[3,1]}}^{11a57} & =&\sigma_{_{[1]}}^4\frac{M_1}{8 A^{8}},\\
_{_4}\sigma_{_{[2,2]}}^{11a57}&=&\sigma_{_{[1]}}^4\frac{M_2}{12 A^{8}}	\\
&&\\
&&\\
_{_4}\sigma_{_{[3]}}^{11a231}& =& _{_4}\sigma_{_{[3]}}^{11a57}-\sigma_{_{[1]}}^5\frac{12 \left(A^2-1\right)^7}{A^2}\\
_{_4}\sigma_{_{[1,1,1]}}^{11a231} & = &_{_4}\sigma^{11a57}_{[1,1,1]}+\sigma_{_{[1]}}^5\frac{24 \left(A^2-1\right)^7}{A^2}\\
_{_4}\sigma_{_{[3,1]}}^{11a231} & = &_{_4}\sigma_{_{[3,1]}}^{11a57} + \sigma_{_{[1]}}^5 \frac{12 \left(A^2-1\right)^7}{A^2}\\
_{_4}\sigma_{_{[2,2]}}^{11a231}&=& _{_4}\sigma_{_{[2,2]}}^{11a57}-\sigma_{_{[1]}}^5\frac{32 \left(A^2-1\right)^7}{A^2},	\\
\end{array}
\end{equation}
where
\begin{equation}
\begin{array}{lll}
M_1 &=&344965 A^{24}-2665432 A^{22}+9074107 A^{20}-17770888 A^{18}+22169870 A^{16}-19748036 A^{14}+17699848 A^{12}-\\ &&22242606 A^{10}+27435212 A^8-23877936 A^6+13148872
A^4-4137552 A^2+569344,\\
M_2 &=& 710559 A^{24}-5402308 A^{22}+17459160 A^{20}-29724688 A^{18}+24275721 A^{16}+1314912 A^{14}-18557728 A^{12}+ \\&&3172952 A^{10}+27157852 A^8-37819168 A^6+24477024
A^4-8223936 A^2+1160344.\\
\end{array}
\end{equation}

\paragraph{$11n71 - 11n75$}

The normalized difference is

\begin{dmath*}
	\Delta H_{[3,1]}^{11n71-11n75} =\{q\}^8 \,[4]^2\,[2]\,D_4 \,D_3\,D_0\,D_{-2}\,A^{-24}\,q^{-46} \left(A^{10} q^{70}-3 A^{10} q^{68}-A^8 q^{68}+3 A^{10} q^{66}+4 A^8 q^{66}-A^{12} q^{64}-A^{10} q^{64}-4 A^8 q^{64}-2 A^6 q^{64}+3 A^{12} q^{62}+2 A^8 q^{62}+A^6 q^{62}-2
	A^{12} q^{60}-5 A^{10} q^{60}+2 A^8 q^{60}-A^6 q^{60}-2 A^{12} q^{58}+5 A^{10} q^{58}+5 A^8 q^{58}-2 A^6 q^{58}+4 A^{12} q^{56}-5 A^8 q^{56}-6 A^6 q^{56}+A^4
	q^{56}+A^{12} q^{54}-7 A^{10} q^{54}+7 A^8 q^{54}+4 A^6 q^{54}+4 A^4 q^{54}-5 A^{12} q^{52}-5 A^{10} q^{52}+5 A^8 q^{52}-8 A^6 q^{52}-A^4 q^{52}+4 A^{12}
	q^{50}+11 A^{10} q^{50}+11 A^8 q^{50}-8 A^6 q^{50}+3 A^4 q^{50}+2 A^{12} q^{48}-11 A^{10} q^{48}-5 A^8 q^{48}-4 A^6 q^{48}+8 A^4 q^{48}-A^{12} q^{46}-8 A^{10}
	q^{46}+5 A^8 q^{46}-11 A^6 q^{46}-2 A^2 q^{46}-3 A^{12} q^{44}+3 A^{10} q^{44}+29 A^8 q^{44}-A^6 q^{44}+9 A^4 q^{44}-2 A^2 q^{44}+7 A^{12} q^{42}-2 A^{10}
	q^{42}-5 A^8 q^{42}-28 A^6 q^{42}+6 A^4 q^{42}-A^2 q^{42}-A^{12} q^{40}-6 A^{10} q^{40}+10 A^8 q^{40}-8 A^6 q^{40}+9 A^4 q^{40}-3 A^2 q^{40}-5 A^{12} q^{38}-16
	A^{10} q^{38}+13 A^8 q^{38}-7 A^6 q^{38}+11 A^4 q^{38}-5 A^2 q^{38}+8 A^{12} q^{36}+9 A^{10} q^{36}+26 A^8 q^{36}-15 A^6 q^{36}+11 A^4 q^{36}+q^{36}-2 A^{12}
	q^{34}-10 A^{10} q^{34}-2 A^8 q^{34}-35 A^6 q^{34}+A^4 q^{34}-9 A^2 q^{34}-8 A^{10} q^{32}+15 A^8 q^{32}+A^6 q^{32}+31 A^4 q^{32}-A^2 q^{32}+q^{32}+2 A^{12}
	q^{30}-7 A^{10} q^{30}+10 A^8 q^{30}-22 A^6 q^{30}+3 A^4 q^{30}-5 A^2 q^{30}+q^{30}+A^{12} q^{28}+6 A^{10} q^{28}+32 A^8 q^{28}-14 A^6 q^{28}+5 A^4 q^{28}-9 A^2
	q^{28}-2 A^{12} q^{26}-16 A^{10} q^{26}-11 A^8 q^{26}-27 A^6 q^{26}+22 A^4 q^{26}+3 A^2 q^{26}+2 q^{26}+4 A^{12} q^{24}+A^{10} q^{24}+16 A^8 q^{24}-8 A^6 q^{24}+6
	A^4 q^{24}-9 A^2 q^{24}-2 A^{12} q^{22}-5 A^{10} q^{22}+20 A^8 q^{22}+12 A^4 q^{22}-5 A^2 q^{22}+q^{22}+A^{12} q^{20}-3 A^{10} q^{20}-3 A^8 q^{20}-30 A^6 q^{20}+2
	A^4 q^{20}-2 A^2 q^{20}+q^{20}+2 A^{12} q^{18}+2 A^{10} q^{18}+11 A^8 q^{18}+15 A^4 q^{18}-A^2 q^{18}-2 A^{12} q^{16}-8 A^{10} q^{16}+A^8 q^{16}-8 A^6 q^{16}+4
	A^4 q^{16}-5 A^2 q^{16}+A^{12} q^{14}+4 A^{10} q^{14}+12 A^8 q^{14}-4 A^6 q^{14}+5 A^4 q^{14}+q^{14}-3 A^{10} q^{12}-4 A^8 q^{12}-12 A^6 q^{12}-A^4 q^{12}-2 A^2
	q^{12}+6 A^8 q^{10}+4 A^6 q^{10}+8 A^4 q^{10}-2 A^{10} q^8-2 A^8 q^8-7 A^6 q^8-2 A^4 q^8-2 A^2 q^8+2 A^{10} q^6+5 A^8 q^6+2 A^6 q^6+3 A^4 q^6-A^{10} q^4-3 A^8
	q^4-4 A^6 q^4+2 A^8 q^2+A^6 q^2+A^4 q^2-A^6\right)
\end{dmath*}

Coefficients in the genus expansion are
\begin{equation}
\begin{array}{lll}
\sigma_{_{[1]}}& =&\frac{-3 A^6+11 A^4-9 A^2+2}{A^8} ,  \\ &&\\
_{_1}\sigma_{_{[2]}}& =&\frac{-22 A^{12}+189 A^{10}-596 A^8+859 A^6-611 A^4+208 A^2-27}{A^{16}}, \\ &&\\
_{_2}\sigma_{_{[1]}}&=&\sigma_{_{[1]}}^3 \frac{-5 A^6+18 A^4-10 A^2+1}{A^8}, \\
_{_2}\sigma_{_{[1,1]}}&=&\sigma_{_{[1]}}^2\frac{102 A^{12}-1017 A^{10}+3366 A^8-4985 A^6+3715 A^4-1358 A^2+193}{2 A^{16}}, \\
_{_2}\sigma_{_{[3]}}&=& -\sigma_{_{[1]}} \frac{\left(A^2-1\right)^2 \left(344 A^{14}-5399 A^{12}+26061 A^{10}-57177 A^8+62463 A^6-35363 A^4+9947 A^2-1100\right)}{2 A^{24}}, \\
_{_2}\sigma_{_{[2,2]}}&=& _{_1}\sigma_{_{[2]}}^2,\\
&&\\
_{_3}\sigma_{_{[2]}} & = &- \sigma_{_{[1]}}^4 \frac{958 A^{12}-11045 A^{10}+39556 A^8-60163 A^6+44139 A^4-15600 A^2+2155}{8 A^{16}},\\
_{_3}\sigma_{_{[2,1]}} & = & \sigma_{_{[1]}}^3 \frac{(A^2-1) L_1}{3 A^{24}},\\
_{_3}\sigma_{_{[4]}} & = &- \sigma_{_{[1]}}^2 \frac{(A^2-1)L_2}{6 A^{32}}\\
_{_3}\sigma_{_{[2,1,1]}} & =& (_{_1}\sigma_{_{[2]}}) \cdotp (_{_2}\sigma_{_{[1,1]}})  ,\\
&&\\
_{_4}\sigma_{_{[1]}} & = &- \sigma_{_{[1]}}^7 \frac{-2 A^4+10 A^2-3}{A^6},\\
_{_4}\sigma_{_{[1,1]}} & = & \sigma_{_{[1]}}^6 \frac{471 A^{12}-5655 A^{10}+20674 A^8-31517 A^6+23353 A^4-8324 A^2+1146}{2 A^{16}},\\
_{_4}\sigma_{_{[1,1,1,1]}} & = &3 (_{_2}\sigma_{_{[1,1]}})^2 ,\\

\end{array}
\end{equation}
where
\begin{equation}
\begin{array}{lll}
L_1&=&2601 A^{16}-39320 A^{14}+222085 A^{12}-611432 A^{10}+913323 A^8-780548 A^6+381065 A^4-98690 A^2+10500, \\
L_2&=&12660 A^{22}-284493 A^{20}+2479924 A^{18}-11264618 A^{16}+30345707 A^{14}-51744353 A^{12}+57722000 A^{10}-\\&&42575683 A^8+20534306 A^6-6215584 A^4+1069867 A^2-79797.\\
\end{array}
\end{equation}

The difference between the HOMFLY polynomials of knots $11n71$ and $11n75$ emerges only in the forth order of the genus expansion:
\begin{equation}
\begin{array}{lll}
_{_4}\sigma_{_{[3]}}^{11n71}& =& - \sigma_{_{[1]}}^5\frac{\left(A^2-1\right)^2 \left(15242 A^{14}-304469 A^{12}+1775229 A^{10}-4462140 A^8+5433000 A^6-3416423 A^4+1073708 A^2-133718\right)}{6 A^{24}},\\
_{_4}\sigma_{_{[1,1,1]}}^{11n71} & = & \sigma_{_{[1]}}^5\frac{-19319 A^{18}+332367 A^{16}-2138175 A^{14}+6995687 A^{12}-13143243 A^{10}+15072569 A^8-10741545 A^6+4642451 A^4-1113096 A^2+113456}{6 A^{24}} ,\\
_{_4}\sigma_{_{[3,1]}} & =&\sigma_{_{[1]}}^4\frac{M_1}{8 A^{32}},\\
_{_4}\sigma_{_{[2,2]}}^{11n71}&=&\sigma_{_{[1]}}^4\frac{M_2}{12 A^{32}}	\\
&&\\
&&\\
_{_4}\sigma_{_{[3]}}^{11n75}& =& _{_4}\sigma_{_{[3]}}^{11n71}-\sigma_{_{[1]}}^5\frac{12 \left(A^2-1\right)^7}{A^{20}}\\
_{_4}\sigma_{_{[1,1,1]}}^{11n75} & = &_{_4}\sigma^{11n71}_{[1,1,1]}+\sigma_{_{[1]}}^5\frac{24 \left(A^2-1\right)^7}{A^{20}}\\
_{_4}\sigma_{_{[3,1]}}^{11n75} & = &_{_4}\sigma_{_{[3,1]}}^{11n71} + \sigma_{_{[1]}}^5 \frac{12 \left(A^2-1\right)^7}{A^{20}}\\
_{_4}\sigma_{_{[2,2]}}^{11n75}&=& _{_4}\sigma_{_{[2,2]}}^{11n71}-\sigma_{_{[1]}}^5\frac{32 \left(A^2-1\right)^7}{A^{20}},	\\
\end{array}
\end{equation}
where
\begin{equation}
\begin{array}{lll}
M_1 &=&91088 A^{24}-2309849 A^{22}+22425042 A^{20}-115860958 A^{18}+363735505 A^{16}-743042116 A^{14}+1025725436 A^{12}-\\&&973911821 A^{10}+636322140 A^8-280772446
A^6+79818757 A^4-13180446 A^2+959412,\\
M_2 &=& 202368 A^{24}-4744094 A^{22}+44867157 A^{20}-229705244 A^{18}+718595120 A^{16}-1465272630 A^{14}+2019700681 A^{12}-\\&&1914472278 A^{10}+1248338173 A^8-549522810
A^6+155807644 A^4-25655184 A^2+1861865.\\
\end{array}
\end{equation}

\paragraph{$11n73 - 11n74$}

The normalized difference is

\begin{dmath*}
	\Delta H_{[3,1]}^{11n73-11n74} =\{q\}^8 \,[4]^2\,[2]\,D_4 \,D_3\,D_0\,D_{-2}\,A^{-10}\,q^{-44} \left(A^8 q^{84}-A^8 q^{82}-A^6 q^{82}+2 A^8 q^{80}+A^6 q^{80}-2 A^{10} q^{78}-A^8 q^{78}-3 A^6 q^{78}+A^{10} q^{76}+3 A^8 q^{76}+2 A^6 q^{76}+A^4 q^{76}-A^{10} q^{74}-4
	A^6 q^{74}-A^4 q^{74}+A^{12} q^{72}-2 A^{10} q^{72}+4 A^8 q^{72}-2 A^6 q^{72}+3 A^4 q^{72}-2 A^6 q^{70}-A^{12} q^{68}-2 A^{10} q^{68}+7 A^8 q^{68}-2 A^6
	q^{68}+2 A^4 q^{68}+2 A^{12} q^{66}-6 A^{10} q^{66}+5 A^8 q^{66}-9 A^6 q^{66}+2 A^4 q^{66}-2 A^2 q^{66}+A^{12} q^{64}+5 A^{10} q^{64}+5 A^8 q^{64}-3 A^6
	q^{64}+3 A^4 q^{64}+2 A^2 q^{64}-2 A^{12} q^{62}-10 A^{10} q^{62}+2 A^8 q^{62}-6 A^6 q^{62}+6 A^4 q^{62}-3 A^2 q^{62}+4 A^{12} q^{60}-5 A^{10} q^{60}+15 A^8
	q^{60}-10 A^6 q^{60}+A^4 q^{60}-3 A^2 q^{60}+4 A^{10} q^{58}+5 A^8 q^{58}-7 A^6 q^{58}+10 A^4 q^{58}+5 A^2 q^{58}-2 A^{12} q^{56}-9 A^{10} q^{56}+8 A^8
	q^{56}-18 A^6 q^{56}-10 A^2 q^{56}+q^{56}+5 A^{12} q^{54}-10 A^{10} q^{54}+13 A^8 q^{54}-3 A^6 q^{54}+16 A^4 q^{54}+3 A^2 q^{54}-q^{54}-A^{12} q^{52}+5 A^{10}
	q^{52}+4 A^8 q^{52}-18 A^6 q^{52}+A^4 q^{52}-A^2 q^{52}+q^{52}-A^{12} q^{50}-10 A^{10} q^{50}+27 A^8 q^{50}-8 A^6 q^{50}+5 A^4 q^{50}-12 A^2 q^{50}+2 q^{50}+6
	A^{12} q^{48}-11 A^{10} q^{48}-4 A^8 q^{48}-19 A^6 q^{48}+20 A^4 q^{48}+7 A^2 q^{48}-4 q^{48}-4 A^{12} q^{46}+8 A^{10} q^{46}+15 A^8 q^{46}-16 A^6 q^{46}+2 A^4
	q^{46}-4 A^2 q^{46}+6 q^{46}+2 A^{12} q^{44}-17 A^{10} q^{44}+19 A^8 q^{44}+8 A^4 q^{44}-17 A^2 q^{44}-3 q^{44}+4 A^{12} q^{42}-2 A^{10} q^{42}+2 A^8 q^{42}-19
	A^6 q^{42}+19 A^4 q^{42}+17 A^2 q^{42}-q^{42}-4 A^{12} q^{40}+2 A^{10} q^{40}+14 A^8 q^{40}-31 A^6 q^{40}-9 A^4 q^{40}-15 A^2 q^{40}+8 q^{40}+5 A^{12} q^{38}-13
	A^{10} q^{38}+12 A^8 q^{38}+14 A^6 q^{38}+30 A^4 q^{38}-9 A^2 q^{38}-9 q^{38}-A^{12} q^{36}+A^{10} q^{36}+A^8 q^{36}-26 A^6 q^{36}-A^4 q^{36}+15 A^2 q^{36}+6
	q^{36}+23 A^8 q^{34}-9 A^6 q^{34}-2 A^4 q^{34}-21 A^2 q^{34}+2 q^{34}+2 A^{12} q^{32}-12 A^{10} q^{32}-5 A^8 q^{32}-8 A^6 q^{32}+19 A^4 q^{32}+6 A^2 q^{32}-6
	q^{32}-A^{12} q^{30}+6 A^{10} q^{30}+11 A^8 q^{30}-10 A^6 q^{30}+3 A^2 q^{30}+6 q^{30}+A^{12} q^{28}-7 A^{10} q^{28}+6 A^8 q^{28}-2 A^6 q^{28}+11 A^4 q^{28}-14
	A^2 q^{28}-2 A^{10} q^{26}+2 A^8 q^{26}-10 A^6 q^{26}-6 A^4 q^{26}+2 A^2 q^{26}-4 q^{26}+2 A^{10} q^{24}+6 A^8 q^{24}-6 A^6 q^{24}+11 A^4 q^{24}+7 A^2 q^{24}+5
	q^{24}-5 A^{10} q^{22}+A^8 q^{22}-6 A^6 q^{22}+5 A^4 q^{22}-11 A^2 q^{22}-2 q^{22}+2 A^{10} q^{20}+5 A^8 q^{20}+4 A^6 q^{20}+3 A^2 q^{20}-q^{20}-2 A^{10}
	q^{18}-A^8 q^{18}-10 A^6 q^{18}-3 A^4 q^{18}-A^2 q^{18}+3 q^{18}+5 A^8 q^{16}+A^6 q^{16}+8 A^4 q^{16}-A^2 q^{16}-q^{16}-3 A^8 q^{14}-3 A^6 q^{14}-q^{14}+4 A^8
	q^{12}+2 A^6 q^{12}+A^4 q^{12}-2 A^2 q^{12}+q^{12}-A^8 q^{10}-5 A^6 q^{10}-A^4 q^{10}+A^8 q^8+2 A^6 q^8+3 A^4 q^8+2 A^2 q^8-2 A^6 q^6-A^4 q^6-2 A^2 q^6+A^6
	q^4+A^4 q^4-A^6 q^2-A^4 q^2+A^4\right)
\end{dmath*}

Coefficients in the genus expansion are
\begin{equation}
\begin{array}{lll}
\sigma_{_{[1]}}& =&-\frac{4 A^6-11 A^4+8 A^2-2}{A^4} ,  \\ &&\\
_{_1}\sigma_{_{[2]}}& =&\frac{\left(A^2-1\right) \left(40 A^{10}-100 A^8+49 A^6+39 A^4-43 A^2+11\right)}{A^8}, \\ &&\\
_{_2}\sigma_{_{[1]}}&=&-\sigma_{_{[1]}}^3\frac{4 A^6-15 A^4+12 A^2-3}{A^4}, \\
_{_2}\sigma_{_{[1,1]}}&=&\sigma_{_{[1]}}^2\frac{186 A^{12}-780 A^{10}+1391 A^8-1364 A^6+834 A^4-316 A^2+57}{2 A^8}, \\
_{_2}\sigma_{_{[3]}}&=& -\sigma_{_{[1]}} \frac{\left(A^2-1\right)^2 \left(1304 A^{14}-3856 A^{12}+3374 A^{10}-715 A^8+382 A^6-1221 A^4+862 A^2-188\right)}{2 A^{12}}, \\
_{_2}\sigma_{_{[2,2]}}&=& _{_1}\sigma_{_{[2]}}^2,\\
&&\\
_{_3}\sigma_{_{[2]}} & = & \sigma_{_{[1]}}^4\frac{1960 A^{12}-7068 A^{10}+7197 A^8+1022 A^6-6202 A^4+3902 A^2-811}{8 A^8},\\
_{_3}\sigma_{_{[2,1]}} & = & -\sigma_{_{[1]}}^3 \frac{(A^2-1) L_1}{3 A^{12}},\\
_{_3}\sigma_{_{[4]}} & = &\sigma_{_{[1]}}^2 \frac{(A^2-1)L_2}{6 A^{16}}\\
_{_3}\sigma_{_{[2,1,1]}} & =& (_{_1}\sigma_{_{[2]}}) \cdotp (_{_2}\sigma_{_{[1,1]}})  ,\\
&&\\
_{_4}\sigma_{_{[1]}} & = &- \sigma_{_{[1]}}^7\frac{A^6-7 A^4+6 A^2-1}{A^4},\\
_{_4}\sigma_{_{[1,1]}} & = & \sigma_{_{[1]}}^6 \frac{860 A^{12}-3572 A^{10}+6299 A^8-6505 A^6+4657 A^4-2151 A^2+444}{2 A^8},\\
_{_4}\sigma_{_{[1,1,1,1]}} & = &3 (_{_2}\sigma_{_{[1,1]}})^2 ,\\

\end{array}
\end{equation}
where
\begin{equation}
\begin{array}{lll}
L_1&=&8008 A^{16}-34035 A^{14}+55454 A^{12}-37953 A^{10}-1794 A^8+23441 A^6-19524 A^4+7821 A^2-1318, \\
L_2&=&76864 A^{22}-416624 A^{20}+899584 A^{18}-966984 A^{16}+515824 A^{14}-108182 A^{12}-27394 A^{10}+\\&&108904 A^8-166241 A^6+120491 A^4-41987 A^2+5765.\\
\end{array}
\end{equation}

\begin{equation}
\begin{array}{lll}
_{_4}\sigma_{_{[3]}}^{11n73}& =& - \sigma_{_{[1]}}^5\frac{\left(A^2-1\right)^2 \left(71678 A^{14}-199873 A^{12}+159125 A^{10}-27985 A^8+39961 A^6-95709 A^4+65185 A^2-15038\right)}{6 A^{12}},\\
_{_4}\sigma_{_{[1,1,1]}}^{11n73} & = & \sigma_{_{[1]}}^5\frac{-56372 A^{18}+318761 A^{16}-812199 A^{14}+1256919 A^{12}-1348127 A^{10}+1073031 A^8-641975 A^6+277615 A^4-78107 A^2+10706}{6 A^{12}} ,\\
_{_4}\sigma_{_{[3,1]}} & =&\sigma_{_{[1]}}^4\frac{M_1}{8 A^{16}},\\
_{_4}\sigma_{_{[2,2]}}^{11n73}&=&\sigma_{_{[1]}}^4\frac{M_2}{12 A^{14}}	\\
&&\\
&&\\
_{_4}\sigma_{_{[3]}}^{11n74}& =& _{_4}\sigma_{_{[3]}}^{11n73}-\sigma_{_{[1]}}^5\frac{12 \left(A^2-1\right)^7}{A^{10}}\\
_{_4}\sigma_{_{[1,1,1]}}^{11n74} & = &_{_4}\sigma^{11n73}_{[1,1,1]}+\sigma_{_{[1]}}^5\frac{24 \left(A^2-1\right)^7}{A^{10}}\\
_{_4}\sigma_{_{[3,1]}}^{11n74} & = &_{_4}\sigma_{_{[3,1]}}^{11n73} + \sigma_{_{[1]}}^5 \frac{12 \left(A^2-1\right)^7}{A^{10}}\\
_{_4}\sigma_{_{[2,2]}}^{11n74}&=& _{_4}\sigma_{_{[2,2]}}^{11n73}-\sigma_{_{[1]}}^5\frac{32 \left(A^2-1\right)^7}{A^{10}},	\\
\end{array}
\end{equation}
where
\begin{equation}
\begin{array}{lll}
M_1 &=&569344 A^{24}-3743832 A^{22}+10514704 A^{20}-16351858 A^{18}+15300989 A^{16}-9101754 A^{14}+4624494 A^{12}-\\&&4058058 A^{10}+4326923 A^8-3241646 A^6+1521394 A^4-409152
A^2+48372,\\
M_2 &=& 1160344 A^{24}-7454976 A^{22}+19737628 A^{20}-26483656 A^{18}+15615111 A^{16}+3694468 A^{14}-11347356 A^{12}+\\&&4322244 A^{10}+4200568 A^8-5839116 A^6+3175256
A^4-884428 A^2+104153.\\
\end{array}
\end{equation}

\paragraph{$11n76 - 11n78$}

The normalized difference is

\begin{dmath*}
	\Delta H_{[3,1]}^{11n76-11n78} =\{q\}^8 \,[4]^2\,[2]\,D_4 \,D_3\,D_0\,D_{-2}\,A^{-26}\,q^{-46} \left(-A^{10} q^{74}+2 A^{10} q^{72}+A^8 q^{72}-2 A^{10} q^{70}-3 A^8 q^{70}+A^{12} q^{68}+2 A^{10} q^{68}+3 A^8 q^{68}+A^6 q^{68}-2 A^{12} q^{66}-3 A^{10} q^{66}-2 A^8
	q^{66}-A^6 q^{66}+A^{12} q^{64}+5 A^{10} q^{64}+2 A^8 q^{64}+A^6 q^{64}+A^{12} q^{62}-3 A^{10} q^{62}-7 A^8 q^{62}-A^{12} q^{60}+4 A^8 q^{60}+6 A^6 q^{60}-2
	A^{12} q^{58}+A^{10} q^{58}-3 A^8 q^{58}-3 A^6 q^{58}-2 A^4 q^{58}+4 A^{12} q^{56}+9 A^{10} q^{56}-3 A^8 q^{56}+3 A^6 q^{56}+A^4 q^{56}-4 A^{12} q^{54}-7 A^{10}
	q^{54}-10 A^8 q^{54}+7 A^6 q^{54}-2 A^4 q^{54}+2 A^{10} q^{52}-A^8 q^{52}+4 A^6 q^{52}-5 A^4 q^{52}+11 A^{10} q^{50}+3 A^8 q^{50}+9 A^6 q^{50}-2 A^4
	q^{50}-A^{10} q^{48}-26 A^8 q^{48}-A^6 q^{48}-6 A^4 q^{48}+A^2 q^{48}-2 A^{12} q^{46}+4 A^{10} q^{46}-3 A^8 q^{46}+23 A^6 q^{46}-2 A^4 q^{46}+A^2 q^{46}-2
	A^{12} q^{44}+3 A^{10} q^{44}-8 A^8 q^{44}+11 A^6 q^{44}-11 A^4 q^{44}+A^2 q^{44}+2 A^{12} q^{42}+16 A^{10} q^{42}-8 A^8 q^{42}+11 A^6 q^{42}-7 A^4 q^{42}+5 A^2
	q^{42}-5 A^{12} q^{40}-5 A^{10} q^{40}-31 A^8 q^{40}+7 A^6 q^{40}-12 A^4 q^{40}+2 A^{12} q^{38}+11 A^{10} q^{38}-2 A^8 q^{38}+36 A^6 q^{38}-3 A^4 q^{38}+6 A^2
	q^{38}-3 A^{12} q^{36}+6 A^{10} q^{36}-18 A^8 q^{36}+10 A^6 q^{36}-22 A^4 q^{36}+3 A^2 q^{36}-q^{36}+12 A^{10} q^{34}-12 A^8 q^{34}+16 A^6 q^{34}-13 A^4
	q^{34}+3 A^2 q^{34}-2 A^{12} q^{32}-A^{10} q^{32}-27 A^8 q^{32}+20 A^6 q^{32}-6 A^4 q^{32}+9 A^2 q^{32}-q^{32}+11 A^{10} q^{30}-7 A^8 q^{30}+27 A^6 q^{30}-17
	A^4 q^{30}+2 A^2 q^{30}-q^{30}-2 A^{12} q^{28}+5 A^{10} q^{28}-11 A^8 q^{28}+18 A^6 q^{28}-15 A^4 q^{28}+4 A^2 q^{28}+4 A^{10} q^{26}-21 A^8 q^{26}+7 A^6
	q^{26}-14 A^4 q^{26}+7 A^2 q^{26}-2 q^{26}+7 A^{10} q^{24}-10 A^8 q^{24}+24 A^6 q^{24}-5 A^4 q^{24}+5 A^2 q^{24}-2 A^{12} q^{22}-10 A^8 q^{22}+15 A^6 q^{22}-14
	A^4 q^{22}+A^2 q^{22}-q^{22}+A^{12} q^{20}+7 A^{10} q^{20}-4 A^8 q^{20}+10 A^6 q^{20}-10 A^4 q^{20}+7 A^2 q^{20}-q^{20}-A^{12} q^{18}-14 A^8 q^{18}+4 A^6
	q^{18}-9 A^4 q^{18}+A^2 q^{18}+2 A^{10} q^{16}-2 A^8 q^{16}+17 A^6 q^{16}-A^4 q^{16}+3 A^2 q^{16}-6 A^8 q^{14}+2 A^6 q^{14}-11 A^4 q^{14}+A^2 q^{14}-q^{14}+3
	A^{10} q^{12}+7 A^6 q^{12}+2 A^2 q^{12}-A^{10} q^{10}-7 A^8 q^{10}-4 A^4 q^{10}+A^{10} q^8+2 A^8 q^8+7 A^6 q^8+2 A^2 q^8-2 A^8 q^6-A^6 q^6-3 A^4 q^6+2 A^6
	q^4-A^8 q^2-A^4 q^2+A^6\right)
\end{dmath*}

Coefficients in the genus expansion are
\begin{equation}
\begin{array}{lll}
\sigma_{_{[1]}}& =&\frac{4 A^6-13 A^4+10 A^2-2}{A^8} ,  \\ &&\\
_{_1}\sigma_{_{[2]}}& =&-\frac{\left(A^2-1\right) \left(24 A^{10}-212 A^8+538 A^6-512 A^4+199 A^2-27\right)}{A^{16}}, \\ &&\\
_{_2}\sigma_{_{[1]}}&=&\sigma_{_{[1]}}^3\frac{8 A^6-25 A^4+13 A^2-1}{A^8}, \\
_{_2}\sigma_{_{[1,1]}}&=&\sigma_{_{[1]}}^2\frac{186 A^{12}-1556 A^{10}+4622 A^8-6314 A^6+4361 A^4-1472 A^2+193}{2 A^{16}}, \\
_{_2}\sigma_{_{[3]}}&=& \sigma_{_{[1]}} \frac{\left(A^2-1\right)^2 \left(536 A^{14}-7776 A^{12}+37050 A^{10}-78366 A^8+80784 A^6-42409 A^4+10932 A^2-1100\right)}{2 A^{24}}, \\
_{_2}\sigma_{_{[2,2]}}&=& _{_1}\sigma_{_{[2]}}^2,\\
&&\\
_{_3}\sigma_{_{[2]}} & = & -\sigma_{_{[1]}}^4\frac{1080 A^{12}-15100 A^{10}+53974 A^8-78450 A^6+53599 A^4-17258 A^2+2155}{8 A^{16}},\\
_{_3}\sigma_{_{[2,1]}} & = & -\sigma_{_{[1]}}^3 \frac{(A^2-1) L_1}{3 A^{24}},\\
_{_3}\sigma_{_{[4]}} & =- &\sigma_{_{[1]}}^2 \frac{(A^2-1)L_2}{6 A^{32}}\\
_{_3}\sigma_{_{[2,1,1]}} & =& (_{_1}\sigma_{_{[2]}}) \cdotp (_{_2}\sigma_{_{[1,1]}})  ,\\
&&\\
_{_4}\sigma_{_{[1]}} & = &\sigma_{_{[1]}}^7\frac{5 A^4-19 A^2+6}{A^6},\\
_{_4}\sigma_{_{[1,1]}} & = & \sigma_{_{[1]}}^6\frac{1092 A^{12}-10024 A^{10}+31667 A^8-43459 A^6+29131 A^4-9319 A^2+1146}{2 A^{16}},\\
_{_4}\sigma_{_{[1,1,1,1]}} & = &3 (_{_2}\sigma_{_{[1,1]}})^2 ,\\

\end{array}
\end{equation}
where
\begin{equation}
\begin{array}{lll}
L_1&=&4056 A^{16}-63143 A^{14}+343446 A^{12}-892646 A^{10}+1249574 A^8-994727 A^6+449612 A^4-107324 A^2+10500, \\
L_2&=&20928 A^{22}-468464 A^{20}+4016336 A^{18}-17846224 A^{16}+46653488 A^{14}-76550010 A^{12}+81492379 A^{10}-\\&&56897251 A^8+25772417 A^6-7274223 A^4+1160323 A^2-79797.\\
\end{array}
\end{equation}

\begin{equation}
\begin{array}{lll}
_{_4}\sigma_{_{[3]}}^{11n76}& =& - \sigma_{_{[1]}}^5\frac{\left(A^2-1\right)^2 \left(27590 A^{14}-483519 A^{12}+2704473 A^{10}-6417996 A^8+7229652 A^6-4146292 A^4+1182030 A^2-133718\right)}{6 A^{24}},\\
_{_4}\sigma_{_{[1,1,1]}}^{11n76} & = & \sigma_{_{[1]}}^5\frac{47420 A^{18}-671599 A^{16}+3785747 A^{14}-11185230 A^{12}+19281432 A^{10}-20432863 A^8+13487027 A^6-5399832 A^4+1199536 A^2-113456}{6 A^{32}},\\
_{_4}\sigma_{_{[3,1]}} & =&\sigma_{_{[1]}}^4\frac{M_1}{8 A^{32}},\\
_{_4}\sigma_{_{[2,2]}}^{11n76}&=&\sigma_{_{[1]}}^4\frac{M_2}{12 A^{32}}	\\
&&\\
&&\\
_{_4}\sigma_{_{[3]}}^{11n78}& =& _{_4}\sigma_{_{[3]}}^{11n76}-\sigma_{_{[1]}}^5\frac{12 \left(A^2-1\right)^7}{A^{22}}\\
_{_4}\sigma_{_{[1,1,1]}}^{11n78} & = &_{_4}\sigma^{11n76}_{[1,1,1]}+\sigma_{_{[1]}}^5\frac{24 \left(A^2-1\right)^7}{A^{22}}\\
_{_4}\sigma_{_{[3,1]}}^{11n78} & = &_{_4}\sigma_{_{[3,1]}}^{11n76} + \sigma_{_{[1]}}^5 \frac{12 \left(A^2-1\right)^7}{A^{22}}\\
_{_4}\sigma_{_{[2,2]}}^{11n78}&=& _{_4}\sigma_{_{[2,2]}}^{11n76}-\sigma_{_{[1]}}^5\frac{32 \left(A^2-1\right)^7}{A^{22}},	\\
\end{array}
\end{equation}
where
\begin{equation}
\begin{array}{lll}
M_1 &=&179072 A^{24}-4286600 A^{22}+39845868 A^{20}-197296918 A^{18}+592263398 A^{16}-1153447498 A^{14}+1512881107 A^{12}-\\&&1359694664 A^{10}+837470683 A^8-346895278
A^6+92207878 A^4-14186852 A^2+959412,\\
M_2 &=& 314776 A^{24}-7726848 A^{22}+74243884 A^{20}-376217744 A^{18}+1145586248 A^{16}-2249457508 A^{14}+2962885794 A^{12}-\\&&2666907084 A^{10}+1642050175 A^8-679075832
A^6+180065934 A^4-27622484 A^2+1861865.\\
\end{array}
\end{equation}
\subsection{``Non-pretzel'' mutant knots}
We were not able to restore N-dependence of the ``non-pretzel'' 11-crossing mutant knots because their structure is more complicated than that of the pretzel knots. For the $U_q(sl_3)$ group, the differences are

\begin{dmath*}
	\Delta H_{[3,1]}^{11a19 - 11a25}= -\{q\}^{13}\,[8]^2\,[7]\,[6]\,[4]^2\,[3]\,[2]\,q^{-57}\, \left(q^{48}-q^{46}-3 q^{44}+5 q^{42}-q^{40}-4 q^{38}+11 q^{36}-8 q^{34}-5 q^{32}+18 q^{30}-12 q^{28}+2 q^{26}+13 q^{24}-18 q^{22}+8 q^{20}+7 q^{18}-11 q^{16}+9 q^{14}-2
	q^{12}-5 q^{10}+5 q^8-2 q^6+q^2-1\right),
\end{dmath*}

\begin{dmath*}
	\Delta H_{[3,1]}^{11a24 - 11a26}=-\{q\}^{13}\,[8]^2\,[7]\,[6]\,[4]^2\,[3]\,[2]\,q^{-31}\, \left(q^{48}-q^{46}-2 q^{44}+5 q^{42}-3 q^{40}-2 q^{38}+10 q^{36}-13 q^{34}-q^{32}+13 q^{30}-17 q^{28}+7 q^{26}+6 q^{24}-20 q^{22}+12 q^{20}+q^{18}-9 q^{16}+10 q^{14}-5
	q^{12}-3 q^{10}+5 q^8-2 q^6+q^4+q^2-1\right),
\end{dmath*}

\begin{dmath*}
	\Delta H_{[3,1]}^{11a251 - 11a253}=-\{q\}^{13}\,[8]^2\,[7]\,[6]\,[4]^2\,[3]\,[2]\,q^{-17}\, \left(q^{48}-q^{46}+2 q^{42}-4 q^{40}+4 q^{38}+3 q^{36}-9 q^{34}+9 q^{32}-3 q^{30}-10 q^{28}+18 q^{26}-8 q^{24}-5 q^{22}+15 q^{20}-13 q^{18}+2 q^{16}+11 q^{14}-9 q^{12}+3
	q^{10}+3 q^8-4 q^6+2 q^4+q^2-1\right),
\end{dmath*}

\begin{dmath*}
	\Delta H_{[3,1]}^{11a252 - 11a254}=-\{q\}^{13}\,[8]^2\,[7]\,[6]\,[4]^2\,[3]\,[2]\,q^{-43}\, \left(q^{48}-q^{46}-q^{44}+2 q^{42}-4 q^{40}+2 q^{38}+5 q^{36}-10 q^{34}+6 q^{32}+q^{30}-14 q^{28}+15 q^{26}-3 q^{24}-12 q^{22}+15 q^{20}-10 q^{18}-4 q^{16}+12 q^{14}-8
	q^{12}+4 q^8-4 q^6+q^4+q^2-1\right),
\end{dmath*}

\begin{dmath*}
	\Delta H_{[3,1]}^{11n34 - 11n42}=-\{q\}^{13}\,[8]^2\,[7]\,[6]\,[4]^2\,[3]\,[2]\,q^{-9}\, \left(q^{46}-q^{44}+2 q^{40}-4 q^{38}+2 q^{36}+3 q^{34}-4 q^{32}+6 q^{30}-q^{28}-3 q^{26}+6 q^{24}-4 q^{22}+4 q^{20}+2 q^{18}-5 q^{16}+5 q^{14}-2 q^{12}-2 q^{10}+4 q^8-2
	q^6+q^2-1\right),
\end{dmath*}

\begin{dmath*}
	\Delta H_{[3,1]}^{11n35 - 11n43}=-\{q\}^{13}\,[8]^2\,[7]\,[6]\,[4]^2\,[3]\,[2]\,q^{-89}\, \left(2 q^{22}+2 q^{16}-q^{14}+2 q^{12}+3 q^{10}-2 q^8+q^6-q^2+1\right),
\end{dmath*}

\begin{dmath*}
	\Delta H_{[3,1]}^{11n36 - 11n44}=-\{q\}^{13}\,[8]^2\,[7]\,[6]\,[4]^2\,[3]\,[2]\,q^{-37}\, \left(q^{22}-q^{20}+q^{16}-2 q^{14}+3 q^{12}+2 q^{10}-q^8+2 q^6+2\right),
\end{dmath*}

\begin{dmath*}
	\Delta H_{[3,1]}^{11n39 - 11n45}=-\{q\}^{13}\,[8]^2\,[7]\,[6]\,[4]^2\,[3]\,[2]\,q^{-35}\, \left(q^{46}-q^{44}+2 q^{40}-4 q^{38}+q^{34}-5 q^{32}+4 q^{30}-3 q^{28}-4 q^{26}+4 q^{24}-6 q^{22}+4 q^{20}+3 q^{18}-7 q^{16}+6 q^{14}-2 q^{12}-3 q^{10}+5 q^8-2 q^6+q^2-1\right),
\end{dmath*}

\begin{dmath*}
	\Delta H_{[3,1]}^{11n40 - 11n46}=-\{q\}^{13}\,[8]^2\,[7]\,[6]\,[4]^2\,[3]\,[2]\,q^{-63}\, \left(q^{22}-q^{20}-q^{18}-3 q^{14}+q^{12}-3 q^8-q^4-q^2+1\right),
\end{dmath*}

\begin{dmath*}
		\Delta H_{[3,1]}^{11n41 - 11n47}=-\{q\}^{13}\,[8]^2\,[7]\,[6]\,[4]^2\,[3]\,[2]\,q^{-63}\, \left(q^{22}-q^{20}-q^{18}-3 q^{14}+q^{10}-3 q^8-q^4-q^2+1\right),
\end{dmath*}

\begin{dmath*}
	\Delta H_{[3,1]}^{11n151 - 11n152}=-\{q\}^{13}\,[8]^2\,[7]\,[6]\,[4]^2\,[3]\,[2]\,q^{-63}\, \left(q^{46}-q^{44}+2 q^{40}-3 q^{38}+q^{36}+2 q^{34}-4 q^{32}+5 q^{30}-q^{28}-2 q^{26}+5 q^{24}-5 q^{22}+5 q^{20}+4 q^{18}-6 q^{16}+6 q^{14}-2 q^{12}-3 q^{10}+5 q^8-2
	q^6+q^2-1\right),
\end{dmath*}

\ \\ \ \\

For the $U_q(sl_4)$ group, the differences are

\begin{dmath*}
	\Delta H_{[3,1]}^{11a19 - 11a25}=- \{q\}^{11}\,[8]^2\,[7]\,[4]^3\,[2]^2\,q^{-97}\left(q^{104}+q^{102}-2 q^{100}+3 q^{96}-2 q^{94}+2 q^{92}+7 q^{90}-6 q^{88}-q^{86}+5 q^{84}-4 q^{82}+6 q^{80}+10 q^{78}-8 q^{76}+q^{74}-q^{72}-4 q^{70}+14 q^{66}+2
	q^{64}-23 q^{62}+14 q^{60}-11 q^{58}-10 q^{56}+16 q^{54}+11 q^{52}-21 q^{50}+7 q^{48}+2 q^{46}-22 q^{44}+26 q^{42}+q^{40}-8 q^{38}+6 q^{36}+q^{34}-15 q^{32}+6
	q^{30}+14 q^{28}-10 q^{26}-4 q^{24}+11 q^{22}-9 q^{20}-5 q^{18}+10 q^{16}-6 q^{12}+4 q^{10}-3 q^6+2 q^4+q^2-1\right),
\end{dmath*}

\begin{dmath*}
	\Delta H_{[3,1]}^{11a24 - 11a26}= \{q\}^{11}\,[8]^2\,[7]\,[4]^3\,[2]^2\,q^{-61}\left(q^{104}-3 q^{100}+3 q^{98}+3 q^{96}-7 q^{94}+4 q^{92}+4 q^{90}-14 q^{88}+6 q^{86}+7 q^{84}-9 q^{82}+7 q^{80}+2 q^{78}-10 q^{76}+2 q^{74}+10 q^{72}-2 q^{68}+19
	q^{66}-24 q^{64}-9 q^{62}+18 q^{60}-7 q^{58}-10 q^{56}+17 q^{54}-8 q^{52}-35 q^{50}+24 q^{48}-7 q^{46}+13 q^{42}-q^{40}-12 q^{38}+2 q^{36}+6 q^{34}-2 q^{32}+10
	q^{30}+4 q^{28}-10 q^{26}-3 q^{24}+5 q^{22}-5 q^{20}+2 q^{18}+6 q^{16}-4 q^{14}-4 q^{12}+3 q^{10}-q^8+2 q^4-1\right),
\end{dmath*}

\begin{dmath*}
	\Delta H_{[3,1]}^{11a251 - 11a253}= \{q\}^{11}\,[8]^2\,[7]\,[4]^3\,[2]^2\,q^{-43}\left(q^{104}-q^{100}+q^{98}-2 q^{94}+3 q^{92}+q^{90}-5 q^{88}-q^{86}+2 q^{84}-4 q^{82}+3 q^{80}+5 q^{78}-2 q^{76}-5 q^{74}+2 q^{72}-3 q^{70}+4 q^{68}+5 q^{66}+3 q^{64}-10
	q^{62}+4 q^{58}-20 q^{56}+27 q^{54}-13 q^{50}+7 q^{48}+7 q^{46}-15 q^{44}+11 q^{42}+17 q^{40}-13 q^{38}+5 q^{36}-5 q^{32}+5 q^{28}-2 q^{26}-6 q^{24}+6 q^{22}-6
	q^{20}-5 q^{18}+11 q^{16}-5 q^{14}-3 q^{12}+6 q^{10}-2 q^8-2 q^6+3 q^4-1\right),
\end{dmath*}

\begin{dmath*}
	\Delta H_{[3,1]}^{11a252 - 11a254}=- \{q\}^{11}\,[8]^2\,[7]\,[4]^3\,[2]^2\,q^{-79}\left(q^{104}+q^{102}-3 q^{94}+q^{92}+2 q^{90}-q^{88}-2 q^{86}-q^{84}-7 q^{82}-2 q^{80}+2 q^{78}+2 q^{74}-3 q^{72}-4 q^{70}-6 q^{68}+6 q^{66}+6 q^{62}-2 q^{60}+7 q^{58}-15
	q^{56}+2 q^{54}+17 q^{52}-15 q^{50}+11 q^{48}+4 q^{46}-10 q^{44}-4 q^{42}+11 q^{40}-8 q^{38}+2 q^{36}+8 q^{34}-12 q^{32}+q^{30}+7 q^{28}-6 q^{26}-3 q^{24}+9
	q^{22}-14 q^{18}+14 q^{16}+q^{14}-12 q^{12}+9 q^{10}+q^8-6 q^6+3 q^4+q^2-1\right),
\end{dmath*}

\begin{dmath*}
	\Delta H_{[3,1]}^{11n34 - 11n42}= \{q\}^{11}\,[8]^2\,[7]\,[4]^3\,[2]^2\,q^{-33}\left(q^{102}-q^{98}+2 q^{90}-q^{88}+2 q^{86}-2 q^{84}-5 q^{82}+3 q^{80}+2 q^{78}-6 q^{76}+8 q^{74}-5 q^{72}-6 q^{70}+6 q^{68}-8 q^{66}+5 q^{64}+6 q^{62}-5 q^{60}+q^{58}+7
	q^{56}-7 q^{54}+4 q^{52}+4 q^{50}+3 q^{48}-q^{46}+3 q^{44}-q^{40}-q^{38}+2 q^{36}-3 q^{34}+5 q^{32}-q^{30}-6 q^{28}+6 q^{26}-5 q^{24}-q^{22}+4
	q^{20}-q^{16}+q^{14}-2 q^{12}+q^{10}+q^4-1\right),
\end{dmath*}

\begin{dmath*}
	\Delta H_{[3,1]}^{11n35 - 11n43}=- \{q\}^{11}\,[8]^2\,[7]\,[4]^3\,[2]^2\,q^{-161}\left(2 q^{92}+q^{90}+2 q^{86}+q^{84}-4 q^{82}+2 q^{80}+2 q^{78}-3 q^{76}-4 q^{74}+q^{72}-8 q^{70}-9 q^{68}+3 q^{66}-2 q^{64}-2 q^{62}-2 q^{60}+q^{58}-8 q^{56}+5 q^{54}+4
	q^{52}+3 q^{50}+6 q^{48}+4 q^{46}-10 q^{44}+2 q^{42}+10 q^{40}-6 q^{38}+2 q^{36}+9 q^{34}-5 q^{32}-13 q^{30}+11 q^{28}+q^{26}-4 q^{24}+q^{22}+6 q^{20}-7 q^{18}-3
	q^{16}+4 q^{14}-2 q^{10}+q^8-q^4+1\right),
\end{dmath*}

\begin{dmath*}
	\Delta H_{[3,1]}^{11n36 - 11n44}= \{q\}^{11}\,[8]^2\,[7]\,[4]^3\,[2]^2\,q^{-97}\left(q^{96}-q^{94}-q^{92}+2 q^{90}-2 q^{86}+3 q^{84}-6 q^{80}-2 q^{78}+6 q^{76}-5 q^{74}-q^{72}+5 q^{70}-7 q^{66}+5 q^{64}+10 q^{62}-2 q^{60}+4 q^{58}+5 q^{56}-6 q^{52}+6
	q^{50}-3 q^{48}-q^{46}-3 q^{44}-4 q^{42}-7 q^{40}+2 q^{38}-5 q^{36}-4 q^{34}+2 q^{32}-4 q^{28}+q^{26}+4 q^{24}-5 q^{22}+2 q^{18}-q^{14}+2 q^{12}-q^{10}+q^4+q^2+1\right),
\end{dmath*}

\begin{dmath*}
	\Delta H_{[3,1]}^{11n39 - 11n45}=- \{q\}^{11}\,[8]^2\,[7]\,[4]^3\,[2]^2\,q^{-69}\left(q^{102}+q^{100}+q^{98}+q^{96}-q^{92}+q^{90}-4 q^{88}+q^{84}-7 q^{82}-2 q^{80}+q^{78}-11 q^{76}+5 q^{72}-9 q^{70}+10 q^{68}+2 q^{66}-7 q^{64}+7 q^{62}+2 q^{60}-6
	q^{58}+12 q^{56}+3 q^{54}-3 q^{52}-6 q^{50}+4 q^{48}-7 q^{46}-2 q^{44}+10 q^{42}-3 q^{40}-3 q^{38}-3 q^{34}-4 q^{32}+13 q^{30}-6 q^{28}+3 q^{26}+2 q^{24}-6
	q^{22}-q^{20}+3 q^{18}+q^{16}+q^{14}-3 q^{12}+2 q^{10}-q^8-2 q^6+2 q^4+q^2-1\right),
\end{dmath*}

\begin{dmath*}
	\Delta H_{[3,1]}^{11n40 - 11n46}= \{q\}^{11}\,[8]^2\,[7]\,[4]^3\,[2]^2\,q^{-125}\left(q^{92}-2 q^{90}-q^{88}+2 q^{86}-q^{84}-2 q^{82}+5 q^{80}+2 q^{78}-5 q^{76}+q^{74}+8 q^{72}-7 q^{70}+8 q^{66}-2 q^{64}-13 q^{62}+q^{60}+q^{58}-11 q^{56}+2 q^{54}+5
	q^{52}-4 q^{50}-6 q^{48}+6 q^{46}-q^{44}+10 q^{42}+8 q^{40}+6 q^{38}+q^{36}+6 q^{34}-4 q^{32}+4 q^{30}+7 q^{28}-4 q^{24}-3 q^{22}-2 q^{20}-7
	q^{18}+q^{16}+q^{14}-q^{12}-3 q^{10}-q^6+q^2+1\right),
\end{dmath*}

\begin{dmath*}
	\Delta H_{[3,1]}^{11n41 - 11n47}=- \{q\}^{11}\,[8]^2\,[7]\,[4]^3\,[2]^2\,q^{-133}\,\left(q^{96}-q^{92}-q^{90}-q^{88}-4 q^{86}-2 q^{80}-3 q^{78}+q^{76}+7 q^{70}+5 q^{68}+6 q^{66}+2 q^{64}+7 q^{62}+2 q^{60}+5 q^{58}+2 q^{54}-q^{52}-2 q^{50}-7 q^{48}-3
	q^{46}-q^{44}-7 q^{42}-3 q^{40}+4 q^{38}+q^{36}-7 q^{34}+4 q^{32}+2 q^{30}-q^{28}-q^{26}+9 q^{24}-2 q^{22}-3 q^{20}+q^{16}-2 q^{14}+2 q^{12}-q^6-q^4+1\right),
\end{dmath*}

\begin{dmath*}
	\Delta H_{[3,1]}^{11n151 - 11n152}=- \{q\}^{11}\,[8]^2\,[7]\,[4]^3\,[2]^2\,q^{-105}\left(q^{102}+q^{100}+q^{98}+q^{96}+q^{94}+2 q^{90}-q^{88}+q^{86}+2 q^{84}-2 q^{82}-q^{80}+4 q^{78}-4 q^{76}-q^{74}+6 q^{72}-4 q^{70}+5 q^{68}+3 q^{66}-2 q^{64}+q^{60}-4
	q^{58}+2 q^{56}+3 q^{54}+q^{52}-13 q^{50}+5 q^{48}-3 q^{46}-10 q^{44}+13 q^{42}+2 q^{40}-7 q^{38}+4 q^{36}-9 q^{32}+15 q^{30}-4 q^{28}+5 q^{24}-6 q^{22}-3 q^{20}+3
	q^{18}+q^{16}-2 q^{12}+2 q^{10}-q^8-2 q^6+2 q^4+q^2-1\right).
\end{dmath*}

\section{Morton mutant knot polynomials \label{MorKnots}}

In \cite{Mor2}, H. Morton suggested that there are some mutant knots which are not distinguished by the HOMFLY-PT polynomials in representation $[2,1]$. We have evaluated the polynomials of these knots in representation $[4,2]$ of the $U_q(sl_3)$ group:
\begin{dmath*}
	\mathcal{H}_{[4,2]}^{K(1,3,3,-3,-3)}=q^{-176}\left(2 q^{260}+2 q^{258}-33 q^{256}+34 q^{254}+178 q^{252}-368 q^{250}-377 q^{248}+1634 q^{246}-346 q^{244}-4372 q^{242}+4606 q^{240}+7208 q^{238}-16060 q^{236}-4034
	q^{234}+36039 q^{232}-17354 q^{230}-56286 q^{228}+70226 q^{226}+51653 q^{224}-153296 q^{222}+16712 q^{220}+231322 q^{218}-176128 q^{216}-230524 q^{214}+402206
	q^{212}+69864 q^{210}-595275 q^{208}+275168 q^{206}+609430 q^{204}-721862 q^{202}-333770 q^{200}+1090134 q^{198}-232242 q^{196}-1173822 q^{194}+952727
	q^{192}+827252 q^{190}-1577202 q^{188}-41986 q^{186}+1806129 q^{184}-989298 q^{182}-1415202 q^{180}+1881766 q^{178}+424748 q^{176}-2210770 q^{174}+827248
	q^{172}+1763276 q^{170}-1827329 q^{168}-688476 q^{166}+2158594 q^{164}-582342 q^{162}-1699331 q^{160}+1545580 q^{158}+644296 q^{156}-1820666 q^{154}+571278
	q^{152}+1289220 q^{150}-1410774 q^{148}-187296 q^{146}+1471058 q^{144}-944128 q^{142}-723268 q^{140}+1531430 q^{138}-438059 q^{136}-1294416 q^{134}+1430022
	q^{132}+376580 q^{130}-1804580 q^{128}+790062 q^{126}+1423672 q^{124}-1720608 q^{122}-456044 q^{120}+2054310 q^{118}-717936 q^{116}-1670956 q^{114}+1645228
	q^{112}+739780 q^{110}-1987752 q^{108}+349532 q^{106}+1679918 q^{104}-1182402 q^{102}-934636 q^{100}+1523786 q^{98}+90481 q^{96}-1381578 q^{94}+572164
	q^{92}+925426 q^{90}-916515 q^{88}-365652 q^{86}+936548 q^{84}-120410 q^{82}-714695 q^{80}+418512 q^{78}+383452 q^{76}-498038 q^{74}-78751 q^{72}+411684
	q^{70}-114772 q^{68}-255384 q^{66}+187499 q^{64}+109102 q^{62}-178500 q^{60}-6664 q^{58}+130887 q^{56}-50396 q^{54}-71346 q^{52}+68422 q^{50}+17189 q^{48}-55690
	q^{46}+17148 q^{44}+27498 q^{42}-25937 q^{40}-2652 q^{38}+17084 q^{36}-8174 q^{34}-4927 q^{32}+7230 q^{30}-1564 q^{28}-2712 q^{26}+2279 q^{24}-42 q^{22}-964
	q^{20}+546 q^{18}+80 q^{16}-242 q^{14}+98 q^{12}+28 q^{10}-42 q^8+12 q^6+4 q^4-4 q^2+1\right)
\end{dmath*}	

\begin{dmath*}
	\mathcal{H}_{[4,2]}^{K(1,3,-3,3,-3)}=q^{-176}\left(2 q^{260}+2 q^{258}-33 q^{256}+32 q^{254}+182 q^{252}-364 q^{250}-389 q^{248}+1636 q^{246}-332 q^{244}-4386 q^{242}+4592 q^{240}+7232 q^{238}-16046 q^{236}-4066
	q^{234}+36039 q^{232}-17302 q^{230}-56316 q^{228}+70160 q^{226}+51725 q^{224}-153268 q^{222}+16602 q^{220}+231352 q^{218}-176012 q^{216}-230604 q^{214}+402118
	q^{212}+70004 q^{210}-595225 q^{208}+274978 q^{206}+609462 q^{204}-721654 q^{202}-333934 q^{200}+1089982 q^{198}-231994 q^{196}-1173804 q^{194}+952465
	q^{192}+827364 q^{190}-1576990 q^{188}-42182 q^{186}+1806009 q^{184}-989044 q^{182}-1415192 q^{180}+1881506 q^{178}+424884 q^{176}-2210584 q^{174}+827026
	q^{172}+1763228 q^{170}-1827117 q^{168}-688568 q^{166}+2158456 q^{164}-582218 q^{162}-1699313 q^{160}+1545480 q^{158}+644350 q^{156}-1820616 q^{154}+571214
	q^{152}+1289270 q^{150}-1410720 q^{148}-187396 q^{146}+1471076 q^{144}-944004 q^{142}-723406 q^{140}+1531338 q^{138}-437847 q^{136}-1294464 q^{134}+1429800
	q^{132}+376766 q^{130}-1804444 q^{128}+789802 q^{126}+1423682 q^{124}-1720354 q^{122}-456164 q^{120}+2054114 q^{118}-717724 q^{116}-1670844 q^{114}+1644966
	q^{112}+739798 q^{110}-1987504 q^{108}+349380 q^{106}+1679754 q^{104}-1182194 q^{102}-934604 q^{100}+1523596 q^{98}+90531 q^{96}-1381438 q^{94}+572076
	q^{92}+925346 q^{90}-916399 q^{88}-365622 q^{86}+936438 q^{84}-120382 q^{82}-714623 q^{80}+418446 q^{78}+383422 q^{76}-497986 q^{74}-78751 q^{72}+411652
	q^{70}-114758 q^{68}-255360 q^{66}+187485 q^{64}+109088 q^{62}-178486 q^{60}-6662 q^{58}+130875 q^{56}-50392 q^{54}-71342 q^{52}+68420 q^{50}+17189 q^{48}-55690
	q^{46}+17148 q^{44}+27498 q^{42}-25937 q^{40}-2652 q^{38}+17084 q^{36}-8174 q^{34}-4927 q^{32}+7230 q^{30}-1564 q^{28}-2712 q^{26}+2279 q^{24}-42 q^{22}-964
	q^{20}+546 q^{18}+80 q^{16}-242 q^{14}+98 q^{12}+28 q^{10}-42 q^8+12 q^6+4 q^4-4 q^2+1\right)
\end{dmath*}

\begin{dmath*}
	\mathcal{H}_{[4,2]}^{K(3,3,3,-3,-3)}=q^{-224}\left(q^{360}-4 q^{358}+4 q^{356}+10 q^{354}-34 q^{352}+20 q^{350}+72 q^{348}-138 q^{346}-16 q^{344}+310 q^{342}-244 q^{340}-402 q^{338}+712 q^{336}+372 q^{334}-1532
	q^{332}-88 q^{330}+3262 q^{328}-1654 q^{326}-5720 q^{324}+6986 q^{322}+6213 q^{320}-16046 q^{318}-1710 q^{316}+26244 q^{314}-5631 q^{312}-41046 q^{310}+12314
	q^{308}+81556 q^{306}-36520 q^{304}-178644 q^{302}+146308 q^{300}+334408 q^{298}-465493 q^{296}-468484 q^{294}+1150112 q^{292}+360242 q^{290}-2340225
	q^{288}+433340 q^{286}+4037846 q^{284}-2670544 q^{282}-5807194 q^{280}+7375420 q^{278}+6275234 q^{276}-15363550 q^{274}-2663146 q^{272}+26094820 q^{270}-9072316
	q^{268}-35947130 q^{266}+32644571 q^{264}+36806610 q^{262}-67836284 q^{260}-16898720 q^{258}+105829269 q^{256}-34087626 q^{254}-126833206 q^{252}+115834368
	q^{250}+104537961 q^{248}-208873260 q^{246}-18994612 q^{244}+274285216 q^{242}-127096019 q^{240}-266307492 q^{238}+298827574 q^{236}+154374240 q^{234}-434578153
	q^{232}+55958950 q^{230}+466975202 q^{228}-314192552 q^{226}-352180388 q^{224}+534508418 q^{222}+96603368 q^{220}-624569134 q^{218}+231782467 q^{216}+526789274
	q^{214}-519579704 q^{212}-254387438 q^{210}+653096500 q^{208}-101680452 q^{206}-573703278 q^{204}+406866430 q^{202}+311322091 q^{200}-543598890 q^{198}+25439908
	q^{196}+465334382 q^{194}-296967930 q^{192}-217198918 q^{190}+393237346 q^{188}-83335734 q^{186}-281351288 q^{184}+296349164 q^{182}+21676308 q^{180}-321427448
	q^{178}+256841758 q^{176}+145450302 q^{174}-415812894 q^{172}+149853428 q^{170}+373458623 q^{168}-427868780 q^{166}-141788952 q^{164}+562943090 q^{162}-184027047
	q^{160}-495855304 q^{158}+473735728 q^{156}+253251492 q^{154}-619464075 q^{152}+71702848 q^{150}+576465752 q^{148}-362987516 q^{146}-373520784 q^{144}+528968530
	q^{142}+93009306 q^{140}-534010924 q^{138}+167765164 q^{136}+403253988 q^{134}-335360250 q^{132}-202178336 q^{130}+382316172 q^{128}+4209788 q^{126}-326333954
	q^{124}+137168594 q^{122}+212249810 q^{120}-201307340 q^{118}-88867556 q^{116}+197291808 q^{114}-8557035 q^{112}-151006562 q^{110}+65086952 q^{108}+90678850
	q^{106}-83183526 q^{104}-37075452 q^{102}+74881760 q^{100}+297232 q^{98}-54080147 q^{96}+18518878 q^{94}+31702548 q^{92}-23493018 q^{90}-13965685 q^{88}+20437088
	q^{86}+2798020 q^{84}-14389206 q^{82}+2648316 q^{80}+8498708 q^{78}-4297918 q^{76}-4089610 q^{74}+3955322 q^{72}+1337292 q^{70}-2855370 q^{68}+71930
	q^{66}+1702773 q^{64}-579796 q^{62}-824984 q^{60}+603516 q^{58}+290120 q^{56}-448178 q^{54}-21588 q^{52}+272736 q^{50}-85540 q^{48}-130108 q^{46}+103152
	q^{44}+32976 q^{42}-74561 q^{40}+15712 q^{38}+33814 q^{36}-24968 q^{34}-5116 q^{32}+15026 q^{30}-5184 q^{28}-4236 q^{26}+4434 q^{24}-488 q^{22}-1508 q^{20}+936
	q^{18}+70 q^{16}-342 q^{14}+140 q^{12}+32 q^{10}-50 q^8+14 q^6+4 q^4-4 q^2+1\right)
	\end{dmath*}	

\begin{dmath*}
	\mathcal{H}_{[4,2]}^{K(3,3,-3,3,-3)}=q^{-224}\left(q^{360}-4 q^{358}+4 q^{356}+10 q^{354}-34 q^{352}+20 q^{350}+72 q^{348}-138 q^{346}-16 q^{344}+310 q^{342}-244 q^{340}-402 q^{338}+712 q^{336}+372 q^{334}-1532
	q^{332}-88 q^{330}+3262 q^{328}-1654 q^{326}-5720 q^{324}+6986 q^{322}+6213 q^{320}-16050 q^{318}-1694 q^{316}+26236 q^{314}-5685 q^{312}-40942 q^{310}+12300
	q^{308}+81336 q^{306}-36176 q^{304}-178638 q^{302}+145624 q^{300}+335198 q^{298}-465163 q^{296}-470174 q^{294}+1151194 q^{292}+361906 q^{290}-2343139
	q^{288}+433050 q^{286}+4042176 q^{284}-2672732 q^{282}-5812242 q^{280}+7381556 q^{278}+6279224 q^{276}-15375256 q^{274}-2661586 q^{272}+26110802 q^{270}-9085188
	q^{268}-35960444 q^{266}+32670283 q^{264}+36807406 q^{262}-67868858 q^{260}-16880474 q^{258}+105858333 q^{256}-34125590 q^{254}-126847776 q^{252}+115888014
	q^{250}+104526667 q^{248}-208932038 q^{246}-18948124 q^{244}+274328738 q^{242}-127175563 q^{240}-266311642 q^{238}+298919598 q^{236}+154326598 q^{234}-434651303
	q^{232}+56049902 q^{230}+467003602 q^{228}-314302732 q^{226}-352152610 q^{224}+534608570 q^{222}+96522298 q^{220}-624630076 q^{218}+231898621 q^{216}+526788072
	q^{214}-519696188 q^{212}-254320760 q^{210}+653172258 q^{208}-101787656 q^{206}-573712780 q^{204}+406970878 q^{202}+311270327 q^{200}-543662396 q^{198}+25523362
	q^{196}+465340776 q^{194}-297046330 q^{192}-217154820 q^{190}+393280264 q^{188}-83407332 q^{186}-281341402 q^{184}+296414322 q^{182}+21615798 q^{180}-321450744
	q^{178}+256926136 q^{176}+145412396 q^{174}-415878428 q^{172}+149941582 q^{170}+373469877 q^{168}-427970746 q^{166}-141737046 q^{164}+563018648 q^{162}-184124613
	q^{160}-495878622 q^{158}+473848612 q^{156}+253214828 q^{154}-619559787 q^{152}+71790780 q^{150}+576515518 q^{148}-363101554 q^{146}-373509168 q^{144}+529072640
	q^{142}+92943950 q^{140}-534074386 q^{138}+167857042 q^{136}+403266080 q^{134}-335448126 q^{132}-202147764 q^{130}+382379546 q^{128}+4153932 q^{126}-326364116
	q^{124}+137231296 q^{122}+212247014 q^{120}-201360042 q^{118}-88840398 q^{116}+197323046 q^{114}-8593833 q^{112}-151014648 q^{110}+65119542 q^{108}+90670756
	q^{106}-83204940 q^{104}-37060444 q^{102}+74891744 q^{100}+281966 q^{98}-54081465 q^{96}+18530790 q^{94}+31698584 q^{92}-23500000 q^{90}-13959907 q^{88}+20439454
	q^{86}+2793184 q^{84}-14388740 q^{82}+2651130 q^{80}+8497364 q^{78}-4299080 q^{76}-4088474 q^{74}+3955612 q^{72}+1336582 q^{70}-2855314 q^{68}+72302
	q^{66}+1702627 q^{64}-579934 q^{62}-824870 q^{60}+603534 q^{58}+290070 q^{56}-448168 q^{54}-21578 q^{52}+272732 q^{50}-85540 q^{48}-130108 q^{46}+103152
	q^{44}+32976 q^{42}-74561 q^{40}+15712 q^{38}+33814 q^{36}-24968 q^{34}-5116 q^{32}+15026 q^{30}-5184 q^{28}-4236 q^{26}+4434 q^{24}-488 q^{22}-1508 q^{20}+936
	q^{18}+70 q^{16}-342 q^{14}+140 q^{12}+32 q^{10}-50 q^8+14 q^6+4 q^4-4 q^2+1\right)
	\end{dmath*}

\section{$[4,2]$ differences for 11-crossing mutant knots \label{11kn42}}

Differences between the HOMFLY-PT polynomials of 11 crossing mutant knots in representation $[4,2]$ of the $U_q(sl_4)$ group:

\begin{dmath*}
	H_{[4,2]}^{11a57-11a231} = \{q\}^{11}\,[10]\,[8]\,[5]^2\,[4]\,[2]^4\, q^{-76} \left(q^{240}-q^{238}-4 q^{236}+10 q^{234}-3 q^{232}-23 q^{230}+43 q^{228}-15 q^{226}-63 q^{224}+125 q^{222}-83 q^{220}-93 q^{218}+300 q^{216}-305 q^{214}-79 q^{212}+655
	q^{210}-783 q^{208}-36 q^{206}+1317 q^{204}-1612 q^{202}+6 q^{200}+2355 q^{198}-2883 q^{196}+290 q^{194}+3443 q^{192}-4524 q^{190}+1392 q^{188}+3405 q^{186}-5732
	q^{184}+3796 q^{182}+705 q^{180}-4746 q^{178}+6388 q^{176}-4891 q^{174}+145 q^{172}+6319 q^{170}-10679 q^{168}+8452 q^{166}+1285 q^{164}-12851 q^{162}+16545
	q^{160}-7259 q^{158}-9199 q^{156}+19412 q^{154}-14395 q^{152}-1322 q^{150}+14540 q^{148}-16161 q^{146}+7681 q^{144}+3623 q^{142}-11563 q^{140}+13414 q^{138}-8137
	q^{136}-2517 q^{134}+12603 q^{132}-15037 q^{130}+7258 q^{128}+6034 q^{126}-15382 q^{124}+13490 q^{122}-1585 q^{120}-10826 q^{118}+13367 q^{116}-5067 q^{114}-5045
	q^{112}+8197 q^{110}-4337 q^{108}-513 q^{106}+2197 q^{104}-1944 q^{102}+1745 q^{100}-1257 q^{98}-275 q^{96}+1918 q^{94}-2118 q^{92}+462 q^{90}+1520 q^{88}-2006
	q^{86}+599 q^{84}+1499 q^{82}-2018 q^{80}+325 q^{78}+1822 q^{76}-1909 q^{74}-46 q^{72}+1849 q^{70}-1466 q^{68}-275 q^{66}+1364 q^{64}-854 q^{62}-319 q^{60}+764
	q^{58}-312 q^{56}-293 q^{54}+344 q^{52}-16 q^{50}-249 q^{48}+117 q^{46}+89 q^{44}-149 q^{42}+16 q^{40}+83 q^{38}-40 q^{36}-40 q^{34}+36 q^{32}+25 q^{30}-35
	q^{28}-q^{26}+35 q^{24}-9 q^{22}-15 q^{20}+13 q^{18}+7 q^{16}-10 q^{14}-q^{12}+7 q^{10}-2 q^8-3 q^6+2 q^4-1\right)
\end{dmath*}

\begin{dmath*}
\Delta H_{[4,2]}^{11n71-11n75} =- \{q\}^{11}\,[10]\,[8]\,[5]^2\,[4]\,[2]^4\, q^{-234} \left(q^{206}-3 q^{204}-q^{202}+9 q^{200}-12 q^{198}-q^{196}+15 q^{194}-21 q^{192}+17 q^{190}+3 q^{188}-35 q^{186}+74 q^{184}-34 q^{182}-99 q^{180}+209 q^{178}-73
q^{176}-210 q^{174}+338 q^{172}-65 q^{170}-319 q^{168}+348 q^{166}-44 q^{164}-224 q^{162}+162 q^{160}-50 q^{158}-2 q^{156}-43 q^{154}-96 q^{152}+157 q^{150}-44
q^{148}-248 q^{146}+356 q^{144}-152 q^{142}-255 q^{140}+677 q^{138}-502 q^{136}-222 q^{134}+1196 q^{132}-1242 q^{130}+267 q^{128}+1157 q^{126}-1506 q^{124}+823
q^{122}+385 q^{120}-1141 q^{118}+1207 q^{116}-701 q^{114}-188 q^{112}+1076 q^{110}-1497 q^{108}+871 q^{106}+325 q^{104}-1584 q^{102}+1618 q^{100}-537 q^{98}-1090
q^{96}+1827 q^{94}-1308 q^{92}-138 q^{90}+1211 q^{88}-1361 q^{86}+687 q^{84}+269 q^{82}-796 q^{80}+883 q^{78}-505 q^{76}+23 q^{74}+523 q^{72}-691 q^{70}+525
q^{68}+53 q^{66}-498 q^{64}+494 q^{62}-117 q^{60}-216 q^{58}+225 q^{56}-54 q^{54}-85 q^{52}+23 q^{50}+37 q^{48}-59 q^{46}-6 q^{44}+45 q^{42}-49 q^{40}+22
q^{38}+4 q^{36}-22 q^{34}+33 q^{32}-23 q^{30}+4 q^{28}+24 q^{26}-22 q^{24}+7 q^{22}+9 q^{20}-11 q^{18}+3 q^{16}+q^{14}-q^8+q^6-q^4-q^2+1\right),
\end{dmath*}

\begin{dmath*}
\Delta H_{[4,2]}^{11n73-11n74} = \{q\}^{11}\,[10]\,[8]\,[5]^2\,[4]\,[2]^4\, q^{-142} \left(q^{226}+q^{224}-q^{222}+q^{220}+3 q^{218}-3 q^{216}-3 q^{214}+6 q^{212}-q^{210}-12 q^{208}+5 q^{206}+11 q^{204}-25 q^{202}-16 q^{200}+41 q^{198}-18 q^{196}-66
q^{194}+77 q^{192}+25 q^{190}-124 q^{188}+68 q^{186}+113 q^{184}-143 q^{182}-24 q^{180}+224 q^{178}-79 q^{176}-216 q^{174}+324 q^{172}+55 q^{170}-436 q^{168}+355
q^{166}+240 q^{164}-629 q^{162}+285 q^{160}+429 q^{158}-678 q^{156}+71 q^{154}+569 q^{152}-503 q^{150}-275 q^{148}+572 q^{146}-85 q^{144}-561 q^{142}+244
q^{140}+596 q^{138}-762 q^{136}-266 q^{134}+1239 q^{132}-770 q^{130}-811 q^{128}+1782 q^{126}-856 q^{124}-955 q^{122}+1795 q^{120}-732 q^{118}-866 q^{116}+1329
q^{114}-354 q^{112}-519 q^{110}+309 q^{108}+303 q^{106}-65 q^{104}-868 q^{102}+962 q^{100}+337 q^{98}-1682 q^{96}+1297 q^{94}+546 q^{92}-1905 q^{90}+1311
q^{88}+441 q^{86}-1460 q^{84}+879 q^{82}+344 q^{80}-788 q^{78}+258 q^{76}+273 q^{74}-98 q^{72}-340 q^{70}+248 q^{68}+356 q^{66}-645 q^{64}+163 q^{62}+529
q^{60}-611 q^{58}+23 q^{56}+507 q^{54}-393 q^{52}-122 q^{50}+394 q^{48}-172 q^{46}-187 q^{44}+258 q^{42}-32 q^{40}-161 q^{38}+120 q^{36}+35 q^{34}-94 q^{32}+22
q^{30}+51 q^{28}-39 q^{26}-19 q^{24}+39 q^{22}-11 q^{20}-20 q^{18}+21 q^{16}-q^{14}-12 q^{12}+8 q^{10}+q^8-4 q^6+2 q^4+q^2-1\right),
\end{dmath*}

\begin{dmath*}
\Delta H_{[4,2]}^{11n76-11n78} = \{q\}^{11}\,[10]\,[8]\,[5]^2\,[4]\,[2]^4\, q^{-234} \left(q^{202}-q^{200}-q^{198}+3 q^{196}-4 q^{194}-4 q^{192}+14 q^{190}-12 q^{188}-9 q^{186}+31 q^{184}-26 q^{182}-17 q^{180}+56 q^{178}-36 q^{176}-25 q^{174}+77
q^{172}-26 q^{170}-65 q^{168}+95 q^{166}+24 q^{164}-136 q^{162}+102 q^{160}+85 q^{158}-202 q^{156}+101 q^{154}+92 q^{152}-210 q^{150}+65 q^{148}+47 q^{146}-108
q^{144}-38 q^{142}+7 q^{140}+111 q^{138}-239 q^{136}-8 q^{134}+356 q^{132}-410 q^{130}+54 q^{128}+478 q^{126}-429 q^{124}+131 q^{122}+376 q^{120}-314 q^{118}+187
q^{116}+169 q^{114}-71 q^{112}-12 q^{110}+114 q^{108}+92 q^{106}-293 q^{104}+85 q^{102}+259 q^{100}-547 q^{98}+134 q^{96}+194 q^{94}-402 q^{92}+2 q^{90}+114
q^{88}-114 q^{86}-141 q^{84}+61 q^{82}+181 q^{80}-304 q^{78}+106 q^{76}+335 q^{74}-462 q^{72}+239 q^{70}+299 q^{68}-464 q^{66}+274 q^{64}+202 q^{62}-360
q^{60}+232 q^{58}+18 q^{56}-80 q^{54}+16 q^{52}-12 q^{50}+115 q^{48}-178 q^{46}+14 q^{44}+184 q^{42}-226 q^{40}+27 q^{38}+165 q^{36}-155 q^{34}-36 q^{32}+134
q^{30}-58 q^{28}-69 q^{26}+101 q^{24}-12 q^{22}-62 q^{20}+47 q^{18}+2 q^{16}-25 q^{14}+18 q^{12}+3 q^{10}-8 q^8+2 q^6+q^4-2 q^2+1\right).
\end{dmath*}


\end{document}